\documentclass[a4paper,twocolumn,11pt,unpublished]{quantumarticle}
\pdfoutput=1
\usepackage[utf8]{inputenc}
\usepackage[english]{babel}
\usepackage[T1]{fontenc}
\usepackage{amsmath}
\usepackage{hyperref}
\usepackage{subcaption}
\usepackage{graphicx}
\usepackage{multirow}
\usepackage{makecell}
\usepackage[numbers,sort&compress]{natbib}

\begin{document}
\graphicspath{ {./figures/} }
\title{State preparation in quantum algorithms for fragment-based quantum chemistry}
\author{Ruhee D'Cunha}
\affiliation{Department of Chemistry, The University of Chicago, Chicago, IL, United States}
\author{Matthew Otten}
\affiliation{HRL Laboratories LLC, Malibu, CA, United States}
\author{Matthew R. Hermes}
\affiliation{Department of Chemistry, The University of Chicago, Chicago, IL, United States}
\author{Laura Gagliardi}
\affiliation{Department of Chemistry, The University of Chicago, Chicago, IL, United States}
\affiliation{Argonne National Laboratory, Lemont, IL, United States}
\affiliation{James Franck Institute, Chicago, IL, United States}
\affiliation{Pritzker School of Molecular Engineering, Chicago, IL, United States}
\author{Stephen K. Gray}
\affiliation{Center for Nanoscale Materials, Argonne National Laboratory, Lemont, IL, United States}
\maketitle

\begin{abstract}
State preparation for quantum algorithms is crucial for achieving high accuracy in quantum chemistry and competing with classical algorithms. The localized active space unitary coupled cluster (LAS-UCC) algorithm iteratively loads a fragment-based multireference wave function onto a quantum computer. In this study, we compare two state preparation methods, quantum phase estimation (QPE) and direct initialization (DI), for each fragment. We analyze the impact of QPE parameters, such as the number of ancilla qubits and Trotter steps, on the prepared state. We find a trade-off between the methods, where DI requires fewer resources for smaller fragments, while QPE is more efficient for larger fragments. Our resource estimates highlight the benefits of system fragmentation in state preparation for subsequent quantum chemical calculations. These findings have broad applications for preparing multireference quantum chemical wave functions on quantum circuits, particularly via QPE circuits.
\end{abstract}

\section{Introduction}
In recent years, quantum chemistry has witnessed remarkable progress in quantum computing, driven by advancements in hardware and algorithms~\cite{google_ai_quantum_and_collaborators_hartree-fock_2020, Kandala2017, von_burg_quantum_2021}. Quantum computers offer a notable advantage by leveraging the exponential reduction in required qubits compared to classical bits for storage and manipulation of quantum information, thanks to the inherent quantum mechanical characteristics of qubits. This potential enables the simulation of complex chemical systems that may be impractical to compute on classical computers, at least in theory.

In many cases, the description of complex systems with numerous degenerate electronic states requires the use of multireference methods based on a multiconfigurational wave function.
Examples of such methods are multireference configuration interation (MRCI) \cite{buenker_individualized_1974,buenker_applicability_1978}, multireference perturbation theory \cite{andersson_secondorder_1992,mcdouall_simple_1988}, and the complete active space self-consistent field method (CASSCF) \cite{roos_bjorn_o_complete_1980}. The computational cost of these methods scales exponentially with the number of electrons and orbitals in the active space, making accurate calculations for large systems intractable.

When dealing with chemical systems that consist of multiple fragments exhibiting local strong correlation, while being surrounded by a weakly correlated environment, active space-based fragmentation methods can serve as an alternative to CASSCF. These methods can help reduce the computational cost of the calculation, while still maintaining an accurate description of the individual fragments. 
One such method is the localized active-space self-consistent field (LASSCF) method \cite{Hermes2019,Hermes2020}. One of the limitations of fragment-based methods like LASSCF is the inability to recover correlation between fragments. This drawback can be addressed by introducing entanglement between fragments, as demonstrated in the localized active space state interaction (LASSI) method \cite{pandharkar_localized_2022}, which however, reintroduces the factorial scaling of CASSCF. 
A more efficient way to improve the LASSCF description is to use an inter-fragment correlator implemented on a quantum computer. 
Quantum computers are particularly suitable for simulating multireference wave functions due to the compact representation and manipulation of vectors containing multiple electronic configurations on a quantum register. To accurately represent complex chemical systems, multireference states must first be prepared on a quantum computer.  Current methods do not rely on classical algorithms to improve the prepared state; instead, they utilize chemical insights to select crucial configurations and simplify the state preparation step \cite{sugisaki_quantum_2019,stair_multireference_2020,tubman_postponing_2018}.

To leverage the capabilities of quantum computers in capturing fragment correlation following a LASSCF calculation, we have developed the localized active-space unitary coupled cluster (LAS-UCC) method \cite{otten_localized_2022}.
The original LAS-UCC method incorporates
quantum phase estimation (QPE) \cite{abrams_quantum_1999, kitaev_quantum_1997} circuits 
to load multireference fragment wave functions, which are subsequently coupled 
with a unitary coupled cluster (UCC) ansatz \cite{kutzelnigg_pair_1977, Bartlett1988a, hoffmann_unitary_1988}. 
The variational quantum eigensolver (VQE) \cite{Peruzzo2014} is used to iteratively determine the UCC parameters and minimize the energy. This involves loading all fragment wave functions at the beginning of each VQE iteration containing a UCC circuit.
In principle  QPE can be utilized for high-fidelity state preparation, by employing  additional (ancilla) qubits than those required to represent the wave function. The measurement of the ancilla qubits induces a collapse of the system register to an eigenstate of the relevant Hamiltonian 
applied to the system via controlled unitary operations \cite{nielsen_chuang_2010}. 
The loading of a converged multireference wave function utilizes state-of-the-art classical methods to capture strong correlation within a fragment, which is challenging to achieve on a quantum computer with an ansatz due to gate depth and optimization issues. Consequently, the reliable preparation of a state that accurately reproduces the LASSCF wave function on a quantum circuit  is an important consideration in refining the LAS-UCC algorithm.

In this study, we present comprehensive resource and error estimates for LAS-UCC by directly compiling quantum circuits for noise-free quantum devices. We investigate two distinct schemes of state preparation to load the LASSCF wave function onto the quantum circuit prior to conducting a VQE iteration. First, we use QPE in the fragments to prepare the ground state using a fragment Hamiltonian derived from the LASSCF calculation (defined in Section 2.2 below). Second, we directly initialize the circuit with the converged LASSCF wave function using one- and two-qubit gates. We validate our code using real chemical systems that demonstrate the impact of increasing fragment numbers and the level of strong correlation within and between fragments. Additionally, we define a threshold number of qubits that distinguishes regions where it is more cost-effective on a quantum computer to perform initial state loading through DI versus fragmented QPE. While our primary focus is state preparation for the LAS-UCC algorithm, our results offer insights into any QPE-based algorithm, as effective state preparation techniques are vital for the success of QPE~\cite{nielsen_chuang_2010}.

The paper is structured as follows:

Section 2 provides the theoretical background of the LAS-UCC algorithm, along with a description of the state preparation circuits and computational details.

Section 3 presents LAS-UCC results using both methods of state preparation, an analysis of the QPE-based state preparation, resource estimates, and a study on spin states of a transition metal complex using LAS-UCC.

Lastly, Section 4 includes a discussion of the obtained results and concluding remarks.

\section{Theoretical Background}
\subsection{LASSCF}

The LASSCF \cite{Hermes2019, Hermes2020} method is a classical fragment-based approach that incorporates user-selected fragment active spaces, along with treating the inactive space and inter-fragment interactions at a mean-field level, such as the restricted Hartree-Fock method (RHF). \cite{Szabo1996}.

The wave function is thus an anti-symmetrized product of the $K$-fragment CAS wave functions $|\Psi_{A_K}\rangle$ and the inactive mean-field wave function $|\Phi_{D}\rangle$:
\begin{equation}
|\mathrm{LAS}\rangle = (\bigwedge_{K}|\Psi_{A_{K}}\rangle) \wedge|\Phi_{D}\rangle
\label{las_wfn}
\end{equation}
It is variationally optimized to obtain the LASSCF energy.

LASSCF is a more advantageous starting point for hybrid quantum-classical methods compared to the commonly used Hartree-Fock wave function, as it begins from a mean-field reference and incorporates intra-fragment correlation, with the size of the active space and choice of fragments as tunable parameters.
As a fragmentation method, it provides a clear advantage over other classical ways of state preparation for a quantum algorithm, allowing the fragment-wise loading of the multireference wave function obtained.

\subsection{LAS-UCC}
The LAS-UCC algorithm combines LASSCF with a whole-system VQE enabling the inclusion of both intra- and inter-fragment correlation effects.

The algorithm currently begins by performing a LASSCF calculation to convergence on a classical computer. 
The LASSCF wave function thus obtained can be written as in Eq. (\ref{las_wfn}) as a product of fragment wave functions. The fragment wave functions are individually loaded onto a quantum computer and measurements are made of a circuit comprising the fragment circuits as well as a parameterized ansatz, such as the UCC ansatz \cite{kutzelnigg_pair_1977, Bartlett1988a, hoffmann_unitary_1988}.

\begin{figure*}[htb]
\includegraphics[width=\textwidth]{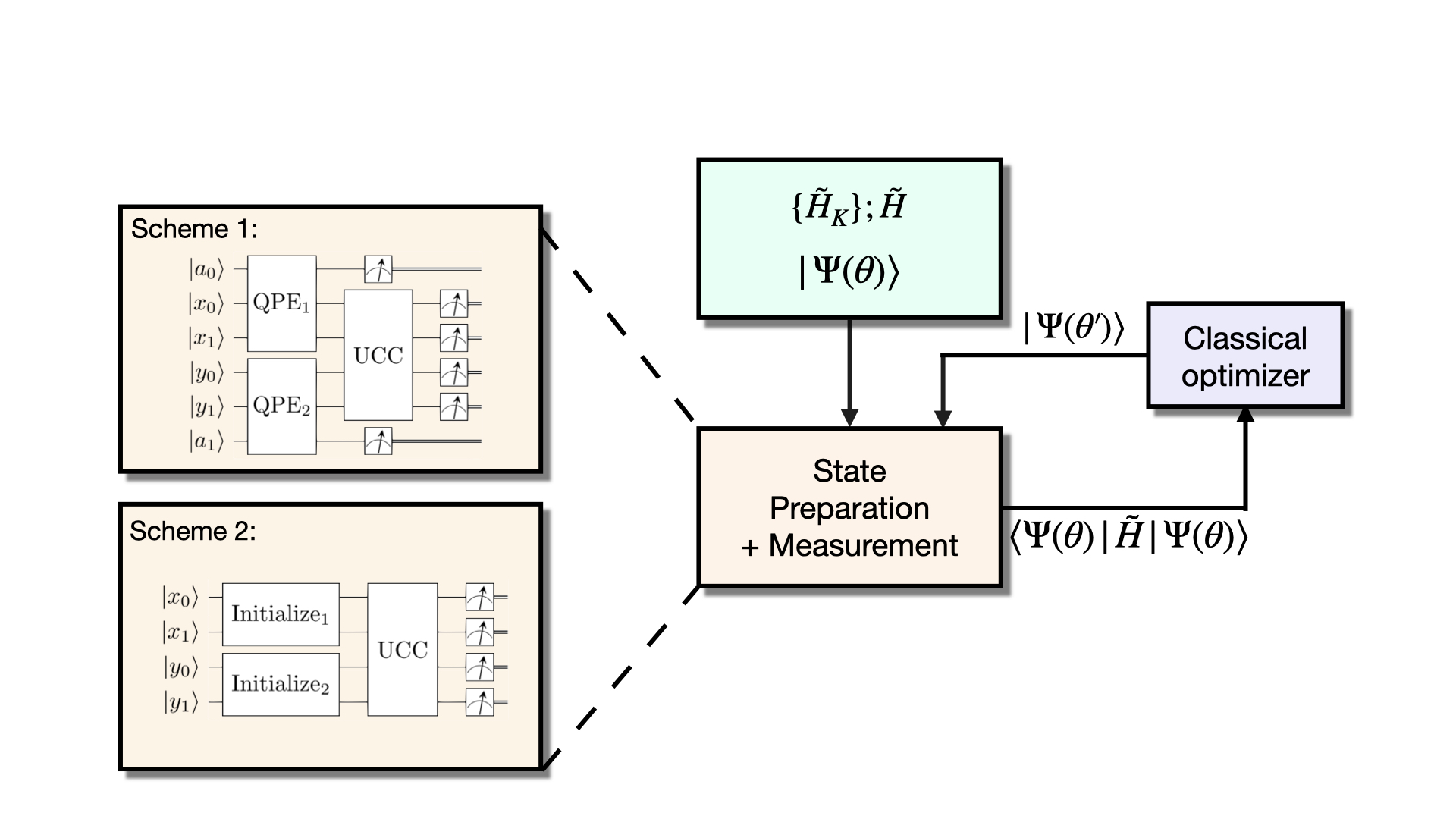}
\caption{Flowchart describing the LAS-UCC algorithm, with example state preparation and measurement circuits containing two 2-qubit fragments, using a single ancilla qubit each for the QPEs. Scheme 1 represents the LAS-UCC algorithm from Ref \citenum{otten_localized_2022} and is referred to within as QPE-based LAS-UCC (QPE-LAS-UCC), and Scheme 2 as direct initialization LAS-UCC (DI-LAS-UCC). The boxes marked `QPE', `Initialize' and `UCC' represent circuits containing one- and two-qubit gates that perform the respective operations. The fragment Hamiltonians $\tilde{H}_K$ and the total system Hamiltonian $\tilde{H}$, both in the qubit basis, as well as the parameterized wave function $|\Psi(\theta)\rangle$, are inputs to the UCC circuit. $|a_i\rangle$ are ancilla qubits, $|x_i\rangle$ and $|y_i\rangle$ are qubits belonging to fragment 1 and fragment 2 respectively. The classical optimizer suggests
new parameters for the UCC  circuits and the overall VQE procedure to minimize the coupled fragments' energy.}
\label{vqe_chart}
\end{figure*}

The localized active space of a single fragment $K$, containing spin orbitals denoted by $(i,j,k,l)$, together with the active spaces of the other fragments $L$, with the corresponding spin orbitals denoted by $(m,n)$ is used to create the fragment Hamiltonian $H_K$:
\begin{align}
    H_K = &\sum_{ij} ( h_{ij} + \sum_{u} g_{iu}^{ju} + \sum_{mn} g_{im}^{jn} \gamma_{m}^{n} )\;\hat{a}^{\dagger}_{j}\hat{a}_{i} \\
    &+ \frac{1}{4} \sum_{ijkl} g_{ij}^{kl} \hat{a}^{\dagger}_{k}\hat{a}^{\dagger}_{l}\hat{a}_{j}  \hat{a}_{i} \notag 
\end{align}
where $h_{ij}$ and $g_{ij}^{kl}$ represent the one- and two-particle components of the Hamiltonian, ${u}$ represents the set of inactive orbitals, and $\hat{a}^{\dagger}_{j}$ and $\hat{a}_{i}$ are fermionic second-quantized creation and annihilation operators. The qubit Hamiltonian $\tilde{H}_K$ is created by mapping the fermionic Hamiltonian $H_K$ to the qubit space via a fermion-to-spin transformation, such as the Jordan-Wigner transformation~\cite{fradkin_jordan-wigner_1989}.

Figure \ref{vqe_chart} provides a flowchart for the quantum portion of the algorithm, once the LASSCF is converged and all conversions to the qubit basis have taken place. The fragment Hamiltonians $\tilde{H}_K$ are used to load the fragment wave functions via QPE circuits, while the total system Hamiltonian $\tilde{H}$ and a generalized UCC ansatz $|\Psi(\theta)\rangle$ are used to compute the VQE energy during the optimization process. A classical optimizer is used to generate new ansatz parameters $\theta'$ to improve the energy measured at each iteration. In this flowchart, the state preparation can be done via either a QPE procedure for each fragment (Scheme 1 in Fig. \ref{vqe_chart}) as originally 
suggested 
\cite{otten_localized_2022}, 
or a direct initialization (DI) of fragment state vectors (Scheme 2), with both sets of circuits depicted as implemented in this work. Other methods may also be used to load the wave function, such as loading individual Slater determinants ~\cite{chee2023shallow}, or a state containing a few Slater determinants, which may be chosen via chemical intuition or an efficient selected configuration interaction algorithm~\cite{tubman_postponing_2018,li2018fast}.

\subsection{State Preparation}

We explore the challenges of achieving both high accuracy in wave function parameters and energies, as well as minimizing the number of qubits and circuit depth needed when using a multireference wave function for state preparation. We investigate two distinct methods of state preparation, which are described below, placing particular emphasis on the necessary resources and the resulting error in the obtained LAS-UCC energies.

\subsubsection{Scheme 1: QPE-based State Preparation}
State preparation begins with a QPE circuit performed on each individual fragment.
The unitary operator for the QPE is given by:
\begin{equation}
    \hat{U}_K = e^{i \tilde{H}_K b}~~~.
    \label{qpe_eq}
\end{equation}
A series of gates controlled by the ancilla qubits and incorporating this unitary operator is applied on the fragment qubits in order to retrieve the phase. After an inverse quantum Fourier transform and measuring the ancilla qubits, the phases $\phi_k$ are obtained as values between 0 and 1 by phase kickback. The eigenvalues $E_k$ of the fragment Hamiltonian $\tilde{H}_K$ can then be obtained as:
\begin{equation}
    E_k= \frac{2 \pi \phi_k}{b}~~~.
\end{equation}

The scaling parameter $b$ must be estimated to lead to
a 1:1 mapping of phase and energy eigenvalues.
Measurement of the ancilla qubits leads to the collapse of the system qubits into one of the eigenstates of the fragment Hamiltonian, with a probability dependent on the overlap of the initial state with the specific eigenstate.

Because the initial state is generally not an eigenstate of the fragment Hamiltonian, the  circuit must be run enough times to obtain the ground state energy (and collapse the system qubits into the ground state with measurement) with high probability. The ancilla qubit phase corresponding to the ground state is stored.
Finally, for the execution of the VQE iteration, the QPE circuit for each fragment is run using the fragment Hamiltonians {$\tilde{H}_K$} until the ancilla phase corresponding to the ground state of the fragment Hamiltonian is reproduced.

\subsubsection{Scheme 2: DI-based State Preparation}
DI is a more straightforward method of state preparation, which entails loading the CI vectors of each individual fragment onto fragment circuits of size  $N_{\textrm{frag}}$, where $N_{\textrm{frag}}$ represents the number of spin orbitals in each fragment's active space. This process involves resetting the fragment qubits to $|0\rangle$ and subsequently applying combinations of one- and two-qubit gates. The angles of these gates are determined classically through a recursive algorithm, allowing for precise setup of the desired state vector on the specified qubits.\cite{shende_synthesis_2006}.

DI of a fragment wave function offers the advantage of entangling only the fragment qubits during state preparation, eliminating the need for ancilla qubits. Unlike the QPE-based method, it does not necessitate running the circuit multiple times to achieve the ground state. However, one drawback of initialization circuits is the exponential number of CNOT gates required. Additionally, DI relies on performing the LASSCF calculation on a classical computer. This approach may face limitations if the fragments are too large to be calculated classically, thus posing a challenge in utilizing DI effectively.

\subsection{Computational Details}
Restricted Hartree-Fock (RHF) calculations are run to obtain reference wave functions for LASSCF calculations using the PySCF program \cite{sun_recent_2020}. The LASSCF wave function is obtained as described in Section 2.1 using the MRH code \cite{hermes_matthew_r_mrh_2023}. Complete active space configuration interaction (CASCI) reference values were generated utilizing the same localized orbital space as LASSCF \cite{Gould1989}. The active space for the CASCI included all fragment active spaces. 
The reference curves are used to benchmark the new methods presented in this work. 
Noise-free simulations of the state preparation and measurement circuits were carried out using the Qiskit framework and the Aer state vector simulator \cite{Qiskit}. The matrix exponentials $e^{i \tilde{H}_K b}$ for the fragment QPEs were approximated by replacing the exponentiated sums of Pauli operators by a transformation into the computational basis, a rotation around the z axis, and a back transformation, combined with a Trotterization with $n$ steps, referred to here as Trotter steps \cite{beig_finding_2005}.

\begin{figure}[htb]
\centering
\begin{subfigure}[b]{0.2\textwidth}
\includegraphics[width=\textwidth]{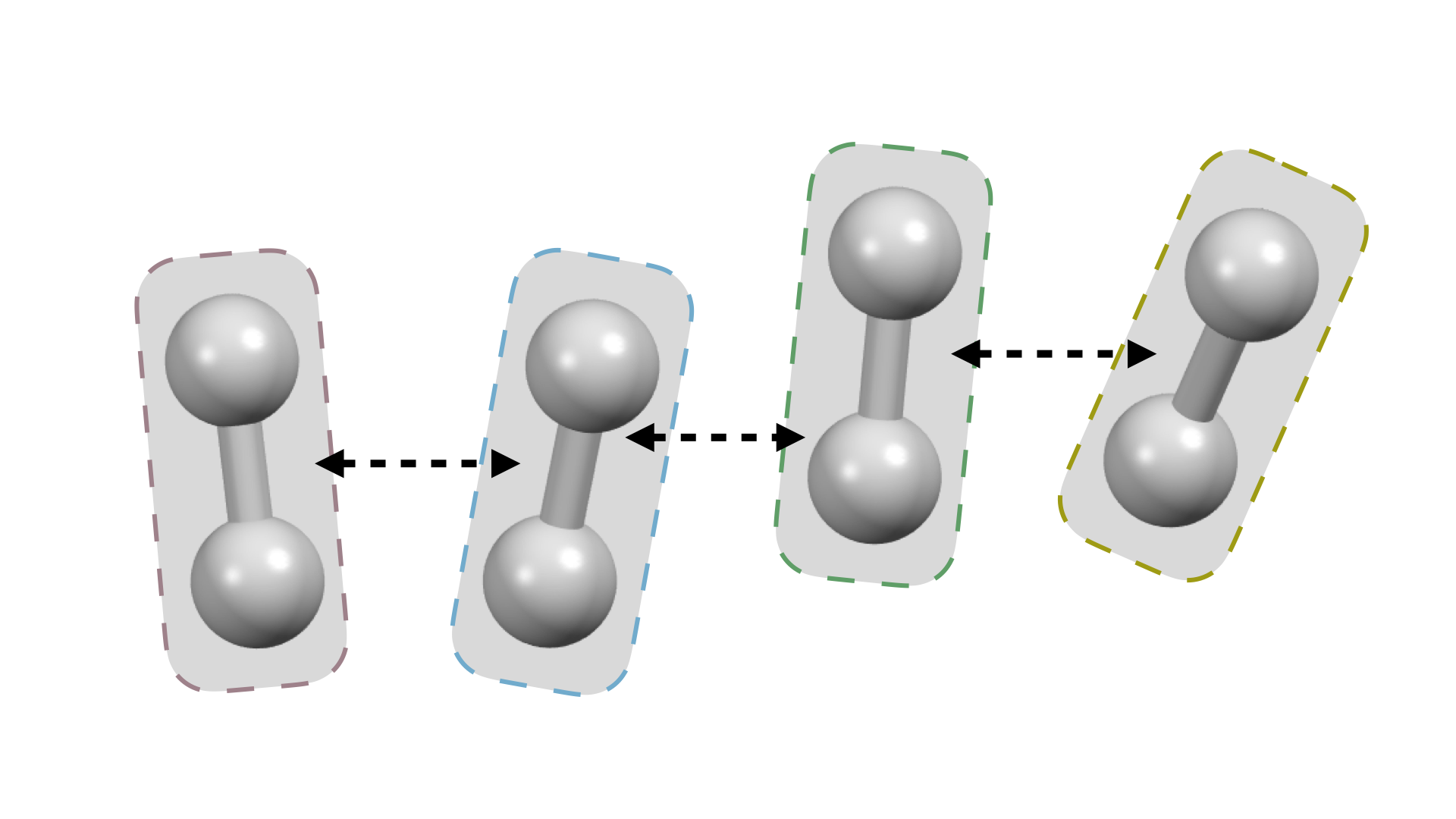}
\caption{}
\end{subfigure}
\begin{subfigure}[b]{0.2\textwidth}
\includegraphics[width=\textwidth]{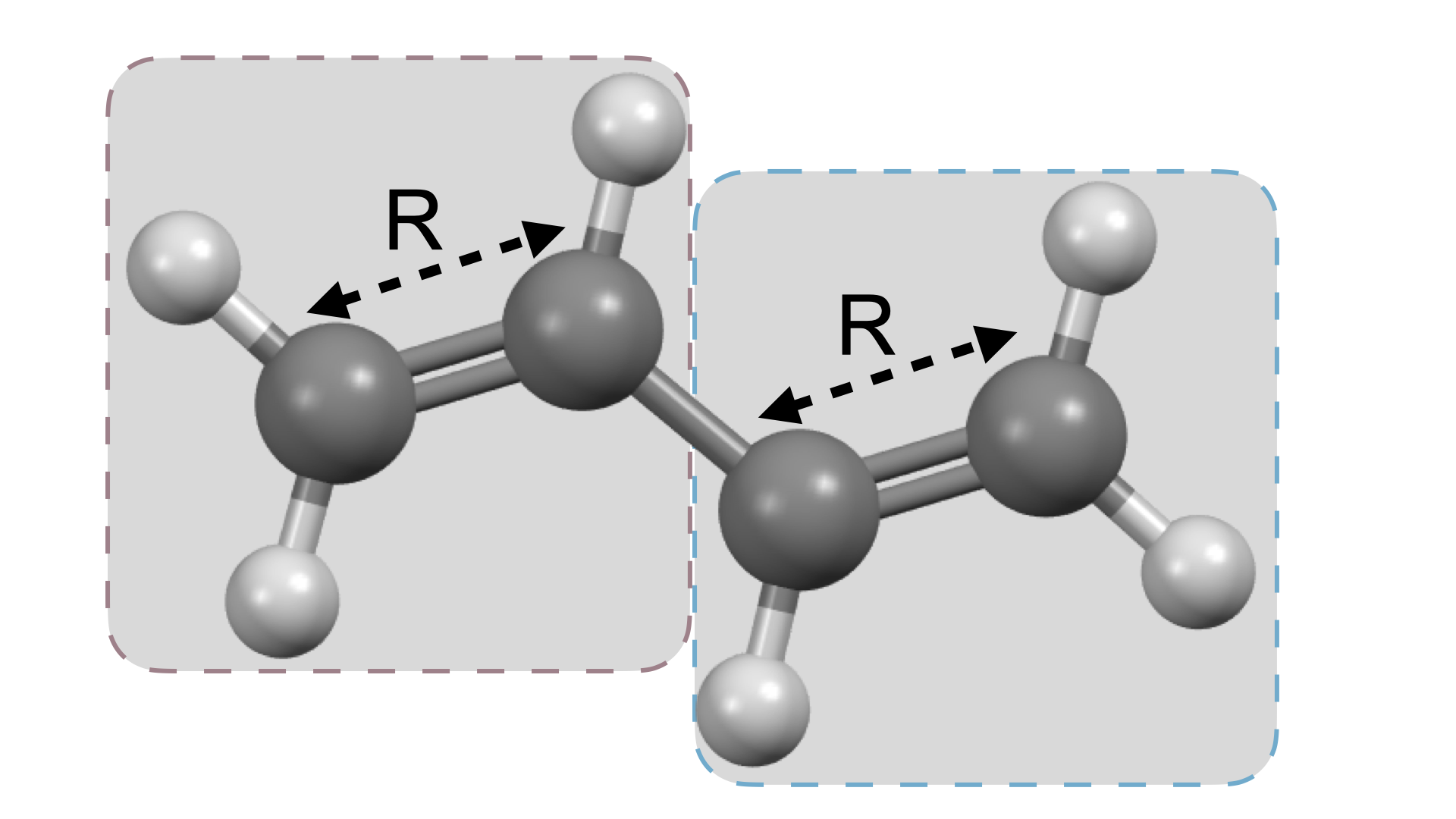}
\caption{}
\end{subfigure}
\begin{subfigure}[b]{0.3\textwidth}
\includegraphics[width=\textwidth]{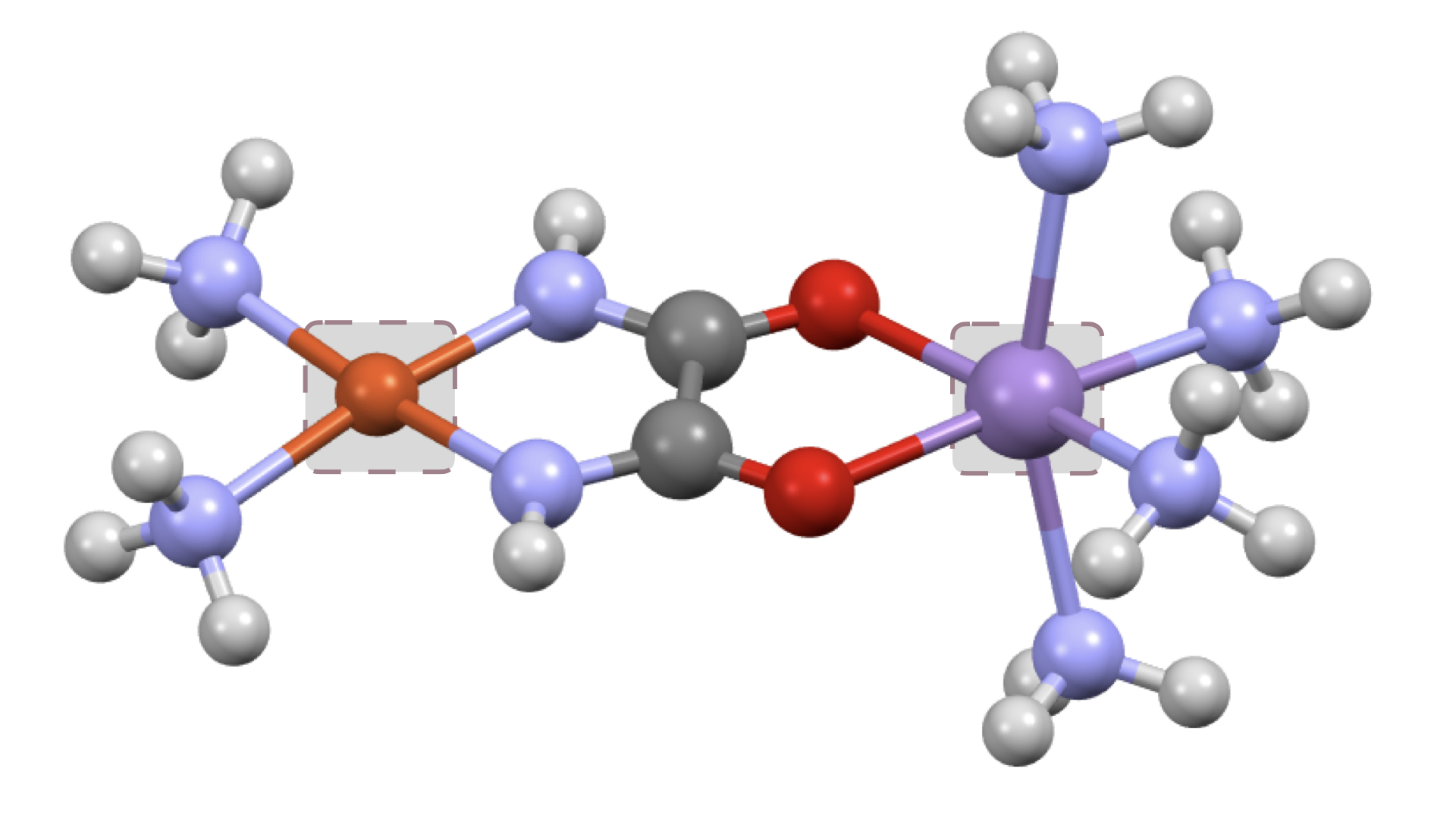}
\caption{}
\label{cumn_sys}
\end{subfigure}
\caption{Systems studied in this work: (a) a set of interacting hydrogen molecules, (b) the trans-butadiene molecule, and (c) a bimetallic complex containing Cu and Mn (Orange: Cu, Purple: Mn, Blue: N, Gray: C). Shaded boxes indicate the fragments used for each system. Arrows represent intermolecular and interatomic distances used to increase or decrease inter- and intra-fragment correlation for the hydrogen and trans-butadiene systems respectively.}
\label{systems}
\end{figure}

The systems studied using LAS-UCC in this work are shown in Figure \ref{systems} and include (a)  a set of interacting hydrogen molecules, in order to study an ideal fragment system with increasing numbers of fragments and the effect of moving the molecules nearer or further away and thus increasing the amount of entanglement between fragments, (b) the trans-butadiene molecule, which is a model system of increasingly stronger correlation within each fragment as the C-C double bonds are simultaneously broken, and (c) a bimetallic system containing copper and manganese, [Mn(NH$_3)_4$]oxamide[Cu(NH$_3)_2]^{2+}$, a transition metal complex with two chemically logical fragments whose spin states are relatively close in energy. An [Fe(H$_2$O)$_4$]$_2$bpym$^{+4}$ (bpym = 2,2’-bipyrimidine) system was used for resource estimations in Section \ref{gate_sec}. Molecular coordinates are provided in the SI.

The fermionic Hamiltonian operator for the VQE in the active space is given by
\begin{equation}
    H_{\textrm{eff}} = \sum_{ij} ( h_{ij} + \sum_{u} g_{iu}^{ju} )\;\hat{a}^{\dagger}_{j}\hat{a}_{i} + \frac{1}{4} \sum_{ijkl} g_{ij}^{kl} \hat{a}^{\dagger}_{k}\hat{a}^{\dagger}_{l}\hat{a}_{j} \hat{a}_{i}.
\end{equation}
This fermionic Hamiltonian $H_{\textrm{eff}}$ is then mapped to a qubit Hamiltonian $\tilde{H}$ using the Jordan-Wigner mapping~\cite{jordan1993paulische}.

The VQE is performed using the circuit needed to load the state vectors as the initial state for each Hamiltonian measurement. The VQE energy is the total (coupled fragments) energy of the system.

\section{Results}

\subsection{Hydrogen systems}
\label{pes}
The potential energy curves for the H$_2$ dimer and trimer
were calculated by varying the separation distance $R$ between the individual H$_2$ molecules at their center of mass. At each geometry, the ground state energy was computed to obtain the respective potential energy curves.

\begin{figure}[htbp]
\includegraphics[width=\linewidth]{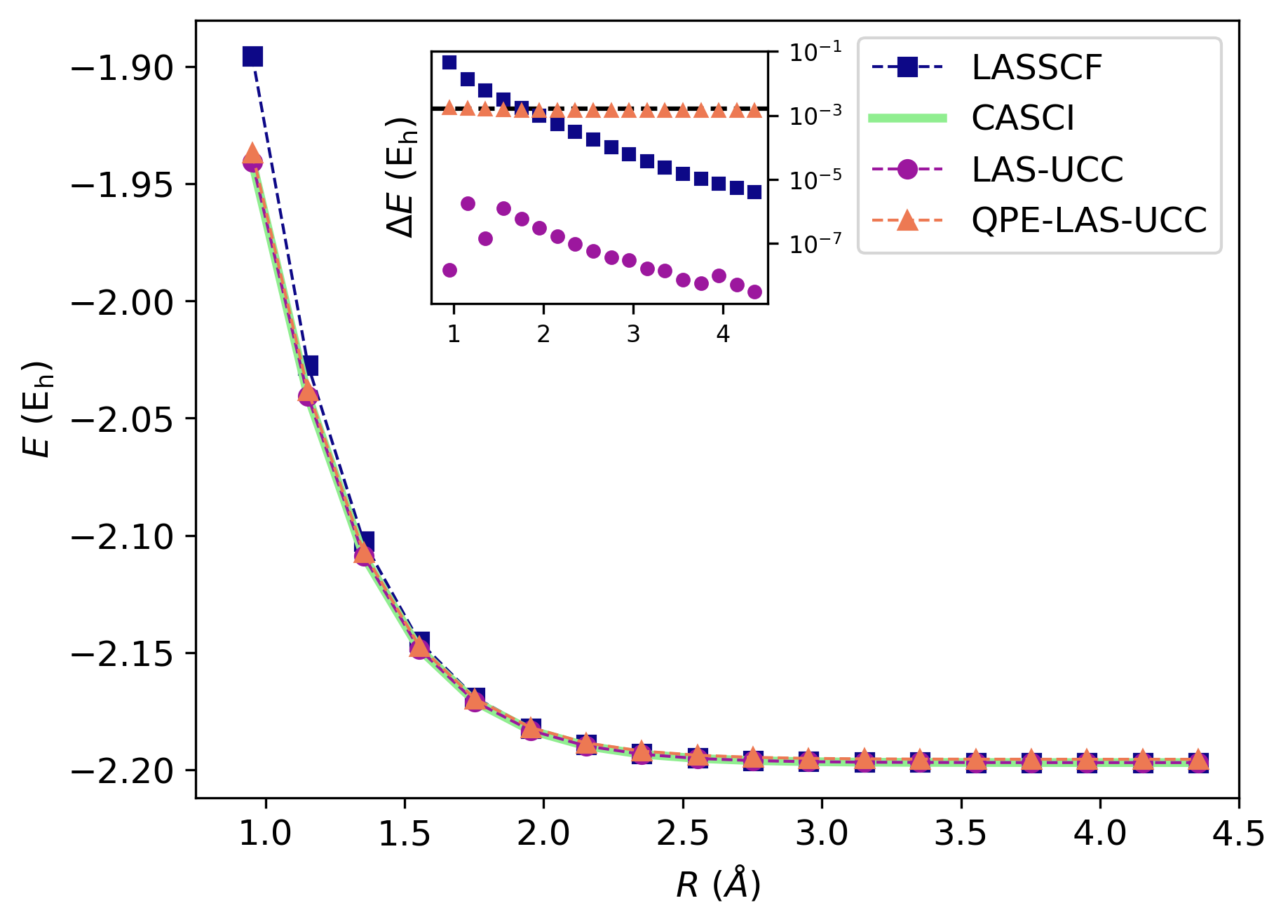}
\caption{VQE energy in Hartrees 
($E_h$) as a function of the distance between molecular centers of mass for (H$_2$)$_2$ as the molecules are moved apart.}
\label{h4_pes}
\end{figure}

The energy curves for (H$_2$)$_2$ obtained using CASCI, LASSCF, the numerically simulated LAS-UCC code, and the QPE-based code (labeled as QPE-LAS-UCC) are depicted in Figure \ref{h4_pes}. Here, by "numerically simulated", we refer to the code that classically minimizes the LAS-UCC energy value with respect to all parameters rather than mapping to a quantum circuit. This approach is a direct measure of the quality of the LAS-UCC algorithm, with no state preparation errors. The QPE-LAS-UCC results are using six Trotter steps and eight ancilla qubits in each fragment. A similar comparison for the (H$_2$)$_3$ system can be found in the SI.

Based on Figure \ref{h4_pes}, it is observed that the numerically simulated LAS-UCC method (blue curve) accurately reproduces the reference curve with high fidelity. On the other hand, the QPE-LAS-UCC method (teal curve) introduces a systematic error across the potential energy curve. However, by utilizing 6 Trotter steps in the QPE-LAS-UCC code, it is possible to achieve an error below chemical accuracy, specifically 1.6 mEh, at all points on the potential energy curve.

The error in the numerically simulated LAS-UCC method roughly follows the error pattern observed in the LASSCF method. It is more significant when the hydrogen molecules are closer together, indicating a less accurate representation of the CASSCF wave function due to stronger interactions between the fragments. As the fragments are pulled apart, the error decreases in magnitude, suggesting an improved accuracy of the LASSCF wave function.

Similarly, the LAS-UCC method, which builds upon the LASSCF wave function, exhibits a similar trend of increased accuracy as the distance between fragments increases. This indicates that the accuracy of the LAS-UCC method also improves with greater separation between the fragments.

The error in the gate-based QPE-LAS-UCC method does not show this trend, as it is induced even at 6 Trotter steps by Trotterization. The magnitude of this error and its dependence on the number of Trotter steps are discussed in Section \ref{trotter_sec} below.

\subsection{State Preparation Method Comparison}
\label{init_sec}
\begin{figure}[ht]
\begin{subfigure}[b]{0.9\columnwidth}
\includegraphics[width=\linewidth]{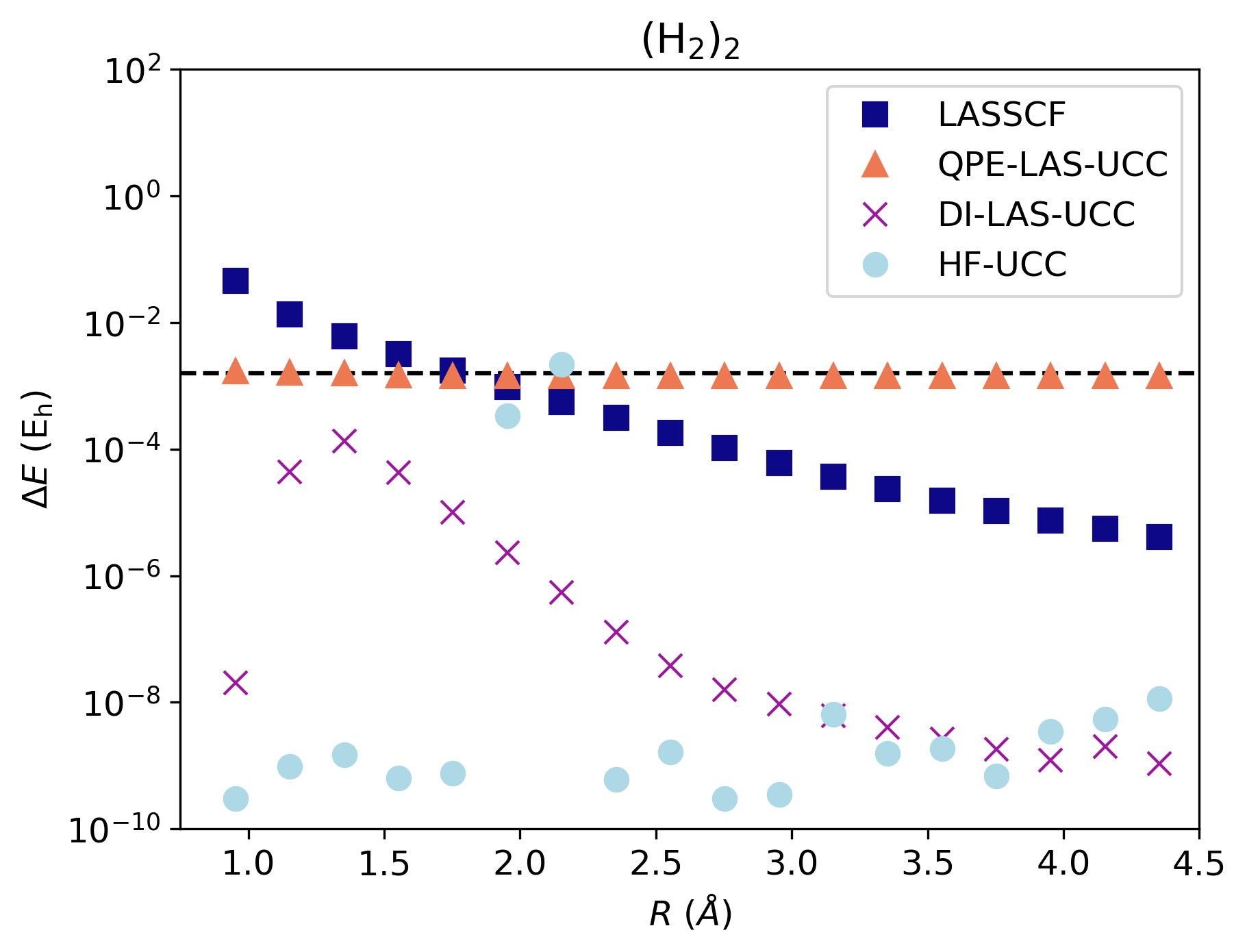}
\caption{}
\label{h4_pes_init}
\end{subfigure}
\begin{subfigure}[b]{0.9\columnwidth}
\includegraphics[width=\linewidth]{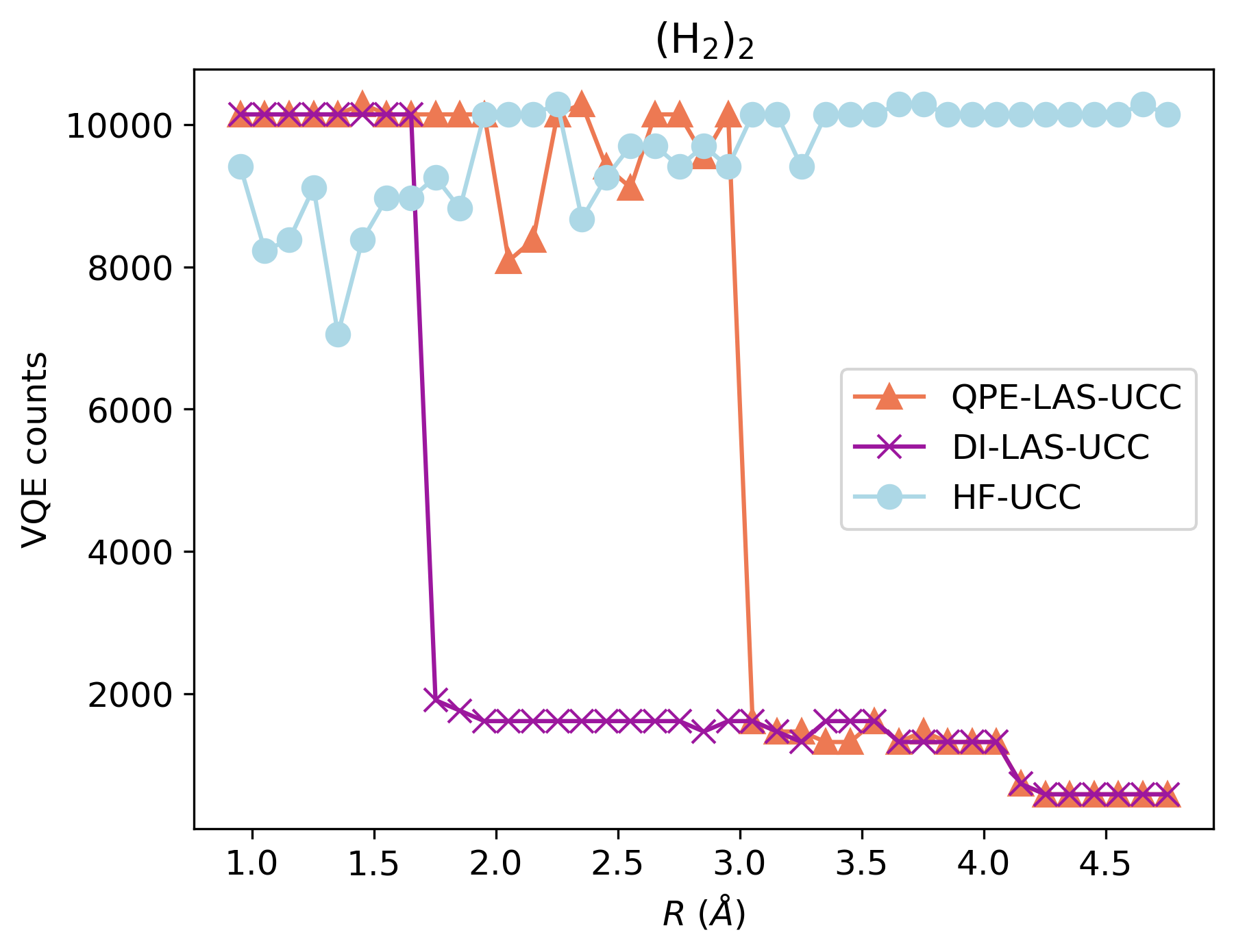}
\caption{}
\label{h4_counts_init}
\end{subfigure}
\caption{(a) VQE energy error with respect to the CASCI reference in Hartrees ($E_h$) and (b) Number of VQE function evaluations, as a function of the distance between molecular centers of mass for (H$_2$)$_2$ as the molecules are moved apart.}
\end{figure}

Figure \ref{h4_pes_init} presents the effects of differing methods of state preparation on the error in the VQE energy for the same potential energy curve as Figure \ref{h4_pes}. Reproduced here are the error with respect to CASCI for classical LASSCF and QPE-LAS-UCC (Inset Fig. \ref{h4_pes}), with the addition of the DI-LAS-UCC and HF-UCC methods. Here, HF-UCC uses a simple Hartree-Fock wave function and the same generalized UCC ansatz as all LAS-UCC methods. Using DI to prepare the state lowers the error significantly as compared to the systematic error shown by the QPE-LAS-UCC method previously.
The energy errors from DI track with the error in the LASSCF energies with respect to the CASCI reference, as expected from a method that eliminates the systematic error associated with Trotterization.
The error with respect to the reference using HF-UCC is also low at most points on the potential energy curve for this small system. However, Figure \ref{h4_counts_init} shows the number of VQE function evaluations (equivalent to VQE iterations, multiplied by a constant system- and optimizer-dependent factor) required for convergence using the Hartree-Fock reference is high for all points on the potential energy curve, while those required for convergence using the DI-LAS-UCC method reduce as the LASSCF wave function becomes a better approximation, beginning at R=1.8 \AA, and remaining low until R=4.5 \AA, the largest separation we studied.

Figure \ref{h4_counts_init} also compares the number of function evaluations required during the VQE optimization for QPE-LAS-UCC and DI-LAS-UCC. DI-LAS-UCC requires fewer function evaluations beginning at $R=1.8$ \AA, while QPE-LAS-UCC requires a higher number until $R=3.0$ \AA. This suggests that the Trotterized state introduces added difficulty to the convergence problem of the VQE that is unrelated to the quality of the LASSCF wave function.

\subsection{QPE-based State Preparation}
The trans-butadiene and interacting H$_2$ model systems have been studied to obtain empirical evidence of the number of qubits and Trotter steps required in fragment QPE calculations for a desired level of accuracy in subsequent QPE-LAS-UCC calculations. The ancilla qubits play an essential role in the identification of the ground state of each fragment, and the Trotter steps affect the fidelity of the loading of the LASSCF wave function onto the circuit, one fragment at a time.

\subsubsection{Number of Ancilla Qubits}
\label{ancilla_sec}
\begin{figure}[ht]
\includegraphics[width=\linewidth]{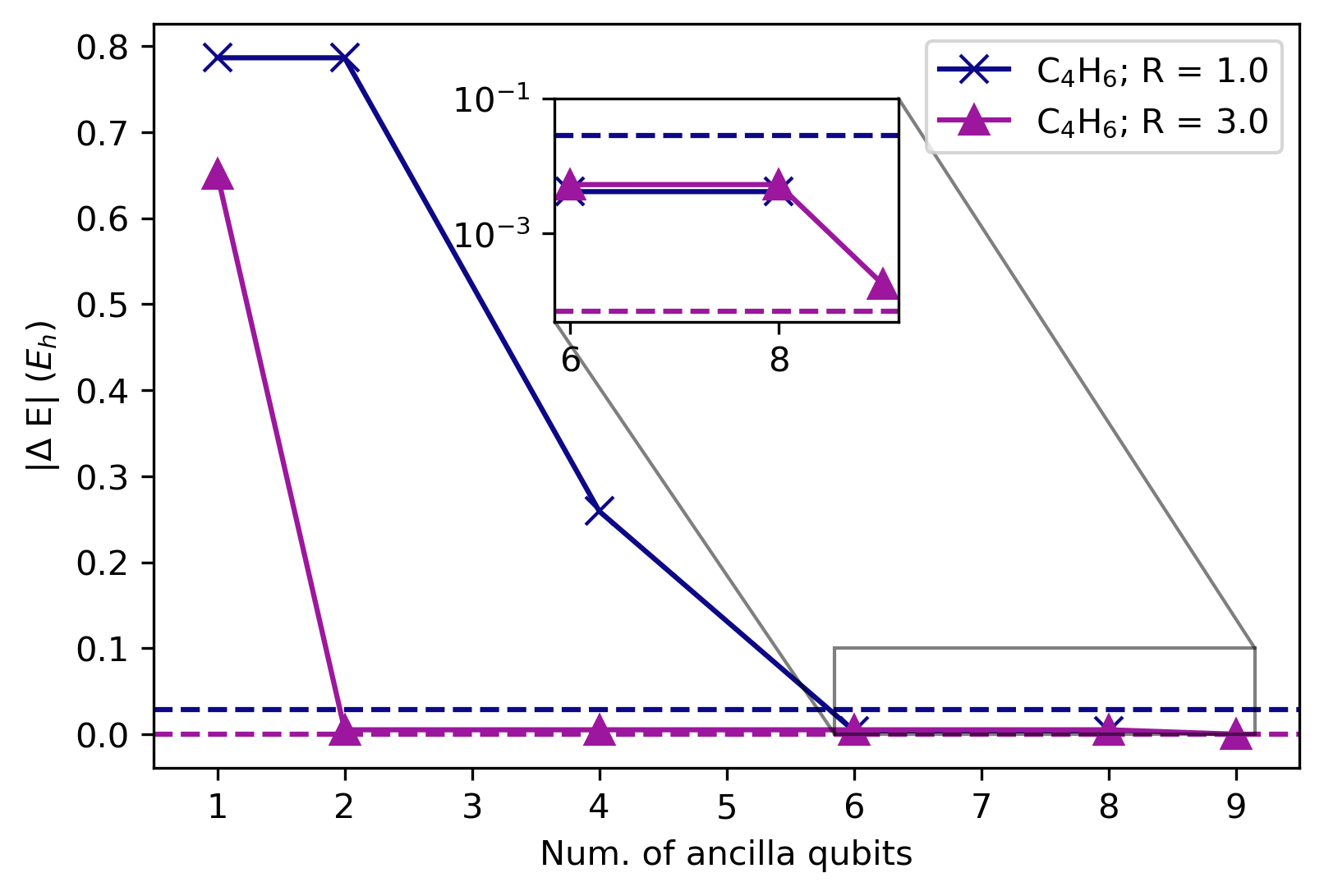}
    \caption{Error in the QPE energy in Hartrees ($E_h$) with respect to the exact diagonalization energy as a function of the number of ancilla qubits used in the QPE for C$_4$H$_6$ with C-C distance = 1.0 (blue) and C$_4$H$_6$ with C-C distance = 3.0 (purple), both with 4 Trotter steps. The corresponding dotted lines represent the gap between the ground and first excited states for each system.}
\label{an_qubits}
\end{figure}

Figure \ref{an_qubits} presents the errors in the QPE energy values of a single fragment for our trans-butadiene system at two different geometries, with R being the C-C double bond distance. This error is calculated with respect to exact diagonalization of the fragment Hamiltonian and is affected by both the number of ancilla qubits as well as the Trotter error. 
Since our chosen fragments have identical geometries and electronic structure, the behavior of the error is identical across fragments, and therefore only the energies corresponding to a single fragment are shown.
As an error threshold we choose the gap between the ground and first excited states, represented by the dotted lines for each geometry. For R=1.0, this threshold is 28.58 m$E_h$, while for R=3.0, it is 0.0071 m$E_h$.

At least six ancilla qubits are required for the error to be smaller than the threshold at R=1.0. 
However, in the strongly-correlated regime where the C-C double bonds of the trans-butadiene molecule are stretched at R=3.0, while the energy error drops significantly at two ancilla qubits, it remains larger than the excitation energy gap threshold. 
While we cannot distinguish between the ground and first excited states of the trans-butadiene system at R=3.0 even with nine ancilla qubits, these states are very close in energy, as can be seen by the purple horizontal dotted line at $10^-5$ in the inset of Fig.\ref{an_qubits}. We note that nine qubits are enough to obtain a sufficiently small energy error, within 1.6 m$E_h$ of the reference energy. Thus, when simulating systems with highly degenerate states, while we cannot guarantee collapse into the ground state with unlimited precision, we can achieve exponential precision with the number of ancilla qubits, though the error must still be minimized with respect to the Trotter steps. 

Further, we observed that in the fragment QPE calculations for trans-butadiene at R=3.0, the most likely eigenvalue for each fragment is no longer the ground state energy (Table in the SI). Further analysis using a Prony-like approach~\cite{hauer1990initial} confirms that the Hartree-Fock reference state has the largest overlap not with the ground state but with an excited state with energy eigenvalue $-2.788 E_h$. 

Another complication of QPE is that the choice of the scale factor, $b$, in Eq. \ref{qpe_eq} can affect the 
precision obtained with a given number of ancillas and some preliminary analysis with differing $b$ values can be beneficial, as we discuss in the SI.
Thus some care and preliminary calculations with post-processing analysis are required to be confident of QPE results in these difficult cases.

On the other hand, the error for a single H$_2$ molecule in the interacting H$_2$ systems, as our ideal systems, requires only more than two ancilla qubits in order to lie below the threshold (Figure in the SI).

\subsubsection{Number of Trotter Steps}
\label{trotter_sec}
\begin{figure}[htbp]
\includegraphics[width=\linewidth]{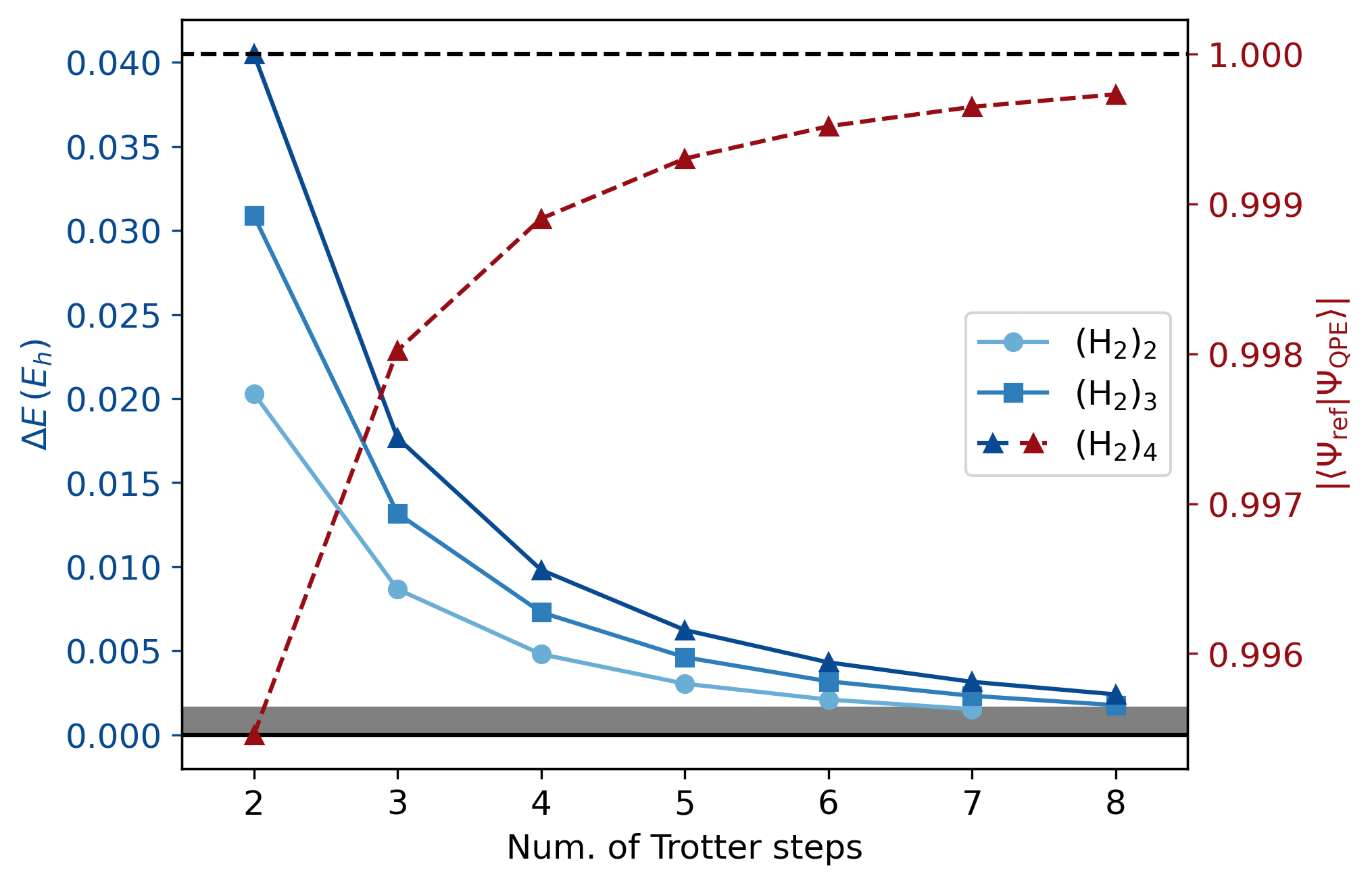}
    \caption{Error in the zeroth-iteration VQE energy with respect to the LASSCF energy in Hartrees (blue) and overlap with the exact diagonalization wave function (red) as a function of the number of Trotter steps used in the QPE for the (H$_2$)$_2$ (light blue/light red), (H$_2$)$_3$ (blue/red), and (H$_2$)$_4$ (dark blue/dark red) systems.}
\label{zeroth_vqe_en}
\end{figure}

Figure \ref{zeroth_vqe_en} contains information about the fidelity of the Trotterized wave function (red) and the energy error for the interacting H$_2$ systems (blue) as a function of the number of Trotter steps. 

The fidelity of the fragment wave function obtained using the Trotter approximation is estimated by the absolute value of its overlap with the one obtained by exact diagonalization. 
The overlaps are averaged over the total number of fragments for the system containing four H$_2$ molecules. The number of ancilla qubits was set to eight (much larger than the two seen to be required in the previous section) to ensure collapse into the ground state, thus allowing us to focus on the effect of increasing Trotter steps only.
As we increase the number of Trotter steps, we see the overlap increase in magnitude, asymptotically approaching 1.

The three blue curves represent the total system VQE energy error at iteration 0, also as a function of the number of Trotter steps. For the H$_2$ tetramer, at two Trotter steps, we see that while the overlap is above 0.995, the VQE energy has an error of 40 m$E_h$ with respect to the LASSCF energy of the total system, taken as a reference. For the H$_2$ dimer, at a minimum 6 Trotter steps are required to converge to within 1.6 m$E_h$ of the reference. The error scales linearly with the number of fragments, thus, the dimer has the lowest error, followed by the trimer and then the tetramer. The per-fragment error remaining constant implies size-intensivity of the method, which is a desirable property.

As the system size increases, a larger number of Trotter steps is then required to converge the 0-th iteration VQE energy to the corresponding LASSCF reference value.
Thus, the fragment QPE wave functions must reproduce the LASSCF wave function with high fidelity in order to minimize the energy error for the whole system. 

\subsection{Resource Estimation}
\label{gate_sec}
The cost of the LAS-UCC algorithm depends on both the cost of the VQE circuit itself and the method chosen for the preparation of the LAS state on the VQE circuit. Assuming the cost of the VQE to be constant, we study the cost of state preparation only in terms of number of gates required for a given accuracy.

In the case of the QPE, the total number of qubits and the gate depth of the final state preparation circuit depend on the number of ancilla qubits and Trotter steps required to model the fragment wave functions with accuracy (in our case within an energy error of 1.6 $mE_h$), which is system-dependent. The DI circuit only needs to be run once and does not require ancilla qubits, however, the gate depth scales exponentially with the size of the active space, with the prefactor depending on the algorithm used to initialize the circuit. DI via quantum multiplexors as implemented in Qiskit scales as $4^N - (3/2) 2^N$, where $N$ is the number of qubits in the fragment~\cite{shende_synthesis_2006}.

\begin{table*}[htbp]
\begin{center}
\begin{tabular}{||c c c c c c||} 
 \hline
 System & \thead{Est. Num. of\\ ancilla qubits} & \thead{Actual\\ ancilla qubits}  & \thead{Num. of\\ Trotter steps} & \thead{CNOT Gates\\ QPE} & \thead{CNOT Gates\\ DI} \\ [0.5ex] 
 \hline\hline
 (H$_2$)$_2$ & 11 & 4 & 7 & 8820 & 232\\ 
 \hline
  (H$_2$)$_4$ & 11 & 4 & 9 & 11340 & 232\\ 
 \hline
  C$_4$H$_6$; R=1.0 & 15 & 6 & 4 & 2,881,872 & 65,152\\ 
 \hline
  C$_4$H$_6$; R=3.0 & 21 & 9 & 4 & 33,231,352 & 65,152\\ 
 \hline
\end{tabular}
\end{center}
\caption{Number of ancilla qubits per fragment estimated for an energy precision of 1.6 $mE_h$. Actual numbers of ancilla qubits used in this study, as well as actual numbers of Trotter steps, with total CNOT gate counts for QPE- and DI-LAS-UCC respectively for each fragment studied.}
\label{an_estimates}
\end{table*}

Table \ref{an_estimates} contains information about the number of ancilla qubits required for a precision of 1.6 $mE_h$ in the fragment QPE energies of each system studied above. This precision threshold was chosen as equivalent to the 1 $kcal mol^{-1}$ threshold of chemical accuracy. The hydrogen systems have the same fragment type, the H$_2$ molecule, and so require the same number of ancilla qubits per fragment for the same precision. The trans-butadiene system, is a model of weak correlation at R=1.0 \AA, requiring 15 ancilla qubits, while at R=3.0 \AA shows more strong correlation and requires 21 qubits per fragment. The details of the ancilla qubit estimation are presented in the SI.
We also report the actual numbers of ancilla qubits used in the simulations for high-accuracy results and the number of CNOT gates required for a single fragment circuit estimated for both QPE- and DI-LAS-UCC. The number of CNOTs required to implement the QPE circuit was computed according to the following equation:
\begin{equation}
    N_{\mathrm{CNOT}} = n_{U} n_{\mathrm{Tr}} (2^{n_{\mathrm{an}}} - 1)
\end{equation}
where $n_{U}$ is the number of CNOTs required to implement a single unitary, $n_{\mathrm{Tr}}$ is the number of Trotter repetitions, and $n_{\mathrm{an}}$ the number of ancilla qubits used in the calculations. This estimation is based on the circuit used for the QPE, with the unitary repeated $n_{\mathrm{Tr}}$ times for the Trotter approximation and $2^{n_{\mathrm{an}}} - 1$ times in a standard QPE circuit.

For the H$_2$ dimer, $N_{\mathrm{CNOT}}$ for QPE-LAS-UCC is equal to 38x $N_{\mathrm{CNOT}}$ for DI-LAS-UCC, while for the tetramer, it is 48x $N_{\mathrm{CNOT}}$ for DI. The number of CNOT gates scales exponentially with the number of ancilla qubits, thus for trans-butadiene at R=3.0 \AA, QPE-LAS-UCC requires an order of magnitude more resources than at R=1.0 \AA. This serves as an example of the system-dependence of the resources required by QPE-LAS-UCC.

To compare the resources needed for state preparation with the two methods, we explored a more realistic chemical system, [Fe(H$_2$O)$_4$]$_2$bpym$^{+4}$ (bpym = 2,2'-bipyrimidine), and studied the effect of increasing the number of active spin orbitals on the total number of CNOT gates required for both methods. Each Fe center was chosen to be a fragment, with the active orbitals being localized on the Fe atoms.

\begin{figure}[htbp]
\includegraphics[width=\linewidth]{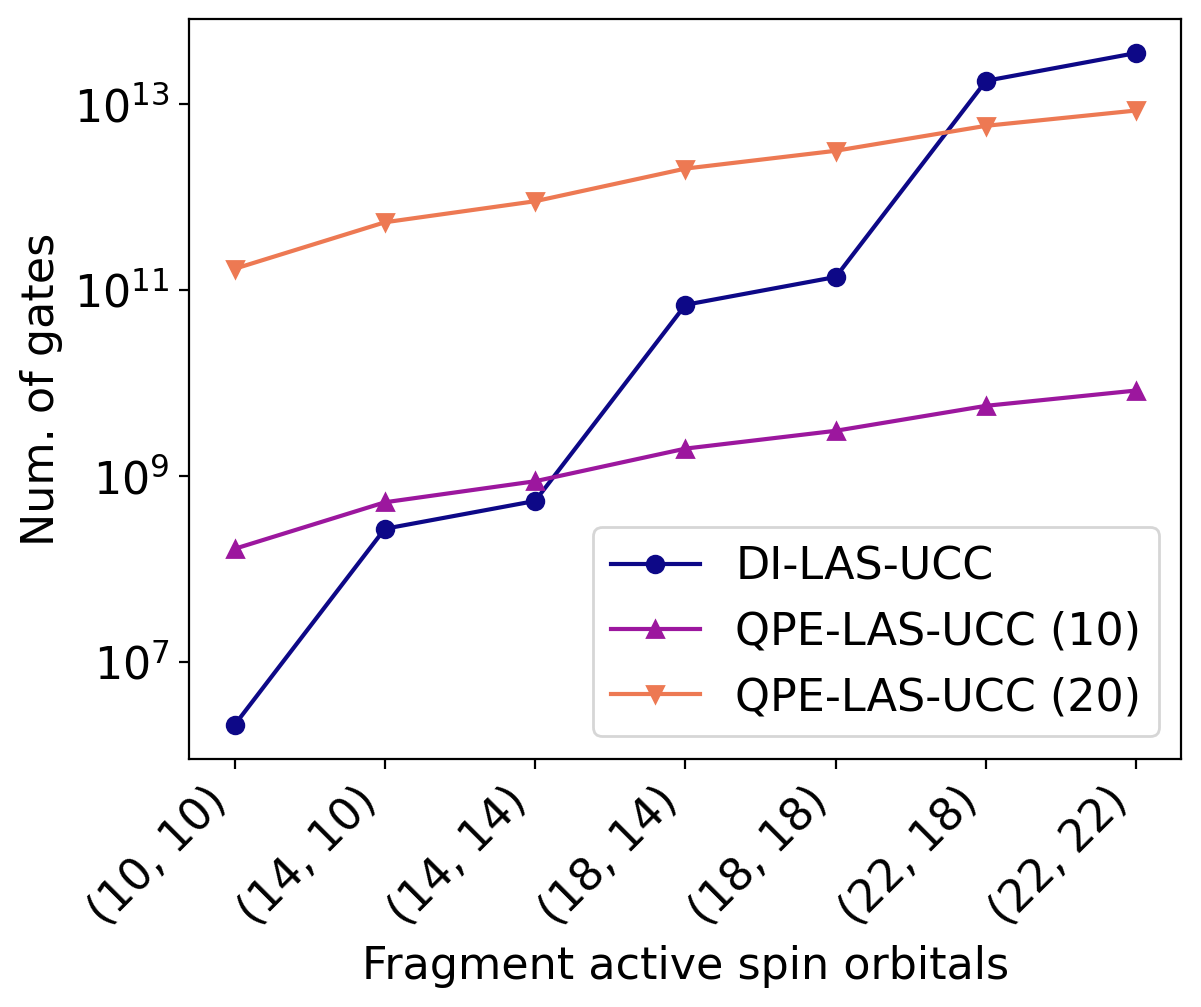}
\caption{Total estimated gate counts for the state preparation circuits tested as a function of the number of spin orbitals in each fragment active space for the [Fe(H$_2$O)$_4$]$_2$bpym$^{+4}$ molecule. QPE-LAS-UCC counts are estimated using 10 Trotter repetitions and 10 (purple) or 20 (peach) ancilla qubits respectively.}
\label{fe_counts}
\end{figure}

Figure \ref{fe_counts} reports the number of CNOT gates estimated for DI-LAS-UCC and QPE-LAS-UCC for different active spaces, ranging from 10 spin orbitals per fragment, with the LAS being ((6,5),(6,5)), to 22 spin orbitals per fragment, with the corresponding LAS as ((12,11),(12,11)). We find on comparing the DI in blue with the two QPE lines in peach and purple that DI requires fewer CNOT gates than QPE for smaller active spaces and cases in which larger numbers of ancilla qubits are used for QPE. As the size of the fragment active spaces grows, QPE-LAS-UCC becomes more efficient, especially if smaller numbers of ancilla qubits are needed. The number of ancilla qubits required, as seen from estimations in Table \ref{an_estimates}, depends on how strongly correlated the individual fragments are and inversely on the gap between the desired ground state and first excited state. (See SI for more details on the ancilla qubit estimates.) For the Fe system, this crossover occurs at between 14 and 20 active fragment spin orbitals. Note that we have here only estimated the gate depth. QPE has additional overheads based on the overlap of the target eigenstate with the initial state. This adds additional overheads in time, which were not estimated here.

\subsection{Application to a Cu-Mn complex}
\label{cumn_sec}

\begin{table*}
\begin{center}
\begin{tabular}{||c c c c c c||} 
 \hline
 $m_s$ & CASSCF & LASSCF & LAS-UCC & DI-LAS-UCC & HF-UCC \\ [0.5ex] 
 \hline\hline
 Septet & 0.0 & 0.0 & 0.0 & 0.0 & 0.0 \\ 
 \hline
 Quintet & -0.04 & 0.01 & -0.03 & -0.03 & -0.03\\
 \hline
 Triplet & 70.58 & 71.51 & 71.48 & 71.48 & 82.47 \\
 \hline
  Singlet & 96.60 & 96.60 & 96.60 & 96.60 & 115.28\\ [1ex] 
 \hline
\end{tabular}
\end{center}
\caption{Energy differences in kcal mol$^{-1}$ for all possible spin states of the Cu-Mn system. CASSCF calculations used a (6,6) active space, while LASSCF and LAS-UCC calculations used ((5,5),(1,1)) fragment active spaces.}
\label{spin_state_gaps}
\end{table*}

Both the numerically simulated LAS-UCC and the DI-LAS-UCC methods were employed for the calculation of spin-state energy differences for a more challenging problem than the model systems above, [Mn(NH$_3)_4$]oxamide[Cu(NH$_3)_2]^{2+}$ (Figure \ref{cumn_sys}). The goal is to compute the spin state energy differences with LAS-UCC and DI-LAS-UCC and compare them with LASSCF, HF-UCC and CASSCF (defined in Section \ref{init_sec}). A minimal (6,6) active space was used for the CASSCF, including 5 $d$ orbitals on the Mn center and 1 $d$ orbital on the Cu center. 
For the LASSCF calculation, the fragment active spaces considered were a (5,5) active space and a (1,1) active space centered on the Mn and Cu atoms respectively.
Table \ref{spin_state_gaps} contains information about the energy differences between states of different $m_s$ values, with antiferromagnetic local spin orientations for the LASSCF subspaces. The LAS-UCC and DI-LAS-UCC methods provide values within 1 kcal mol$^{-1}$ of the CASSCF reference values. However, the HF-UCC method gives an error of close to 10 kcal mol$^{-1}$ for the lower $m_s$ states, which are, in general, harder to simulate. These results suggest that the LAS wave function is a better starting point for the VQE than Hartree-Fock for multi spin-center containing systems, and the LAS-UCC method also improves on the classical LASSCF calculation.

\section{Discussion}

We have analyzed two methods of state preparation for the loading of a fragment multireference wave function onto a quantum circuit, to obtain highly accurate ground state energies of systems with strongly correlated subunits. Section \ref{pes} shows that the error in the QPE-based state preparation (QPE-LAS-UCC) is dominated by Trotter error, which can however be systematically reduced. While both QPE- and DI-LAS-UCC lower the number of VQE iterations as compared to simply using a Hartree-Fock reference (Section \ref{init_sec}), DI provides an ancilla- and probability-free method to load the state, at the cost of exponential scaling in the number of gates. 

The analysis in Sections \ref{ancilla_sec} and \ref{trotter_sec} suggests that the number of ancilla qubits and the number of Trotter steps heavily influence the quality of the fragment QPE wave functions, while the values of these parameters required for a desired accuracy depend on the system size and degree of strong correlation.

Strongly correlated systems, our final target systems, are challenging for the use of QPE for state preparation, requiring 
careful preliminary calculations and post-processing analysis as well as a large number of ancilla qubits and Trotter steps.

For systems with fragments that can be represented by less than 20 qubits, DI-LAS-UCC requires a smaller number of gates than the QPE with 10 Trotter steps.
However, for systems whose representation requires more than 20 qubits, there exists a crossover point where the QPE algorithm's polynomial scaling requires fewer gates than DI. Thus, the size of the active space can guide the choice of state preparation method. These results also provide insight to more general QPE algorithms, where effective state preparation is required, pointing to DI of complex classically computed wave functions as a potential technique for small-scale demonstrations of QPE.

We note that the ancilla-dependence, post-processing requirement, and other overheads will also apply to a full-system QPE, while the fragmentation of the active space results in shallower state-preparation circuits, thus the QPE-LAS-UCC and DI-LAS-UCC have a clear advantage over QPE in terms of the number of gates required.

Our results in Section \ref{cumn_sec} for the bimetallic system compare the LAS-UCC method, simulated both numerically and using a noiseless state vector simulator, with a generalized UCC ansatz utilizing an HF reference. The LAS-UCC method replicates the CASSCF reference value with high accuracy, confirming that to obtain accurate energy differences for spin states of multi-centered transition metal complexes, the LASSCF wave function is an ideal starting point.

Future work includes improvements to the LAS-UCC method through exploration of the VQE ansatz and optimization procedure. Currently the VQE step of the algorithm uses a generalized UCC ansatz, which is physically motivated and accurate, but expensive in terms of gate depth.
Alternative methods of building an ansatz such as ADAPT-VQE \cite{Grimsley2019}, Qubit Coupled Cluster \cite{Ryabinkin2018} or Unitary Selective Coupled Cluster \cite{fedorov_unitary_2022} can be considered in order to reduce the circuit depth. Other more efficient optimization schemes can also be tested \cite{menickelly_latency_2023,kubler_adaptive_2020}.
Our ultimate goal is to simulate complex chemical systems via fragment-based methods by leveraging the power of both classical and quantum computers.

\section{Acknowledgements}
This research is based on work supported by Laboratory Directed Research and Development (LDRD) funding from Argonne National Laboratory, provided by the Director, Office of Science, of the U.S. Department of Energy (DOE) under Contract no. DE-AC0206CH11357. This work was performed, in part, at the Center for Nanoscale Materials, a U.S. Department of Energy Office of Science User Facility, and supported by the U.S. Department of Energy, Office of Science, under Contract no. DE-AC0206CH11357. M.R.H. and L.G. are partially supported by the U.S. DOE, Office of Basic Energy Sciences, Division of Chemical Sciences, Geosciences, and Biosciences under grant no. USDOE/DE-SC002183.  M.J.O. is partially supported by the Defense Advanced Research Projects Agency under Contract No. HR001122C0074.
This material is based upon work supported by the U.S. Department of Energy, Office of Science, National Quantum Information Science Research Centers. We gratefully acknowledge the computing resources provided by University of Chicago Research Computing Center.
\bibliographystyle{quantum}
\bibliography{references}
\end{document}


\title{Supporting Information}

\section*{S1: QPE Scale Factors and Time Series Analysis}

In QPE, some of the eigenvalues, $U_j$,  of a unitary operator $U$ = $e^{iH b}$ are estimated, where $H$ in our case is the fragment Hamiltonian and $b$ is a scaling factor. (We assume $H$ has been written as a sum of Pauli terms and that the overall term proportionate to the identity operator has been subtracted out.) Typically, the eigenvalues are written as $U_j$ = $e^{i 2 \pi \phi_j}$, where $\phi_j$ can be viewed to be in $(0,1)$, or equivalently $(-1/2, 1/2)$ and it is a set of approximate $\phi_j$ that QPE directly estimates. In terms of the Hamiltonian energies we have $E_k = \frac{2 \pi \phi_k}{b}$. The purpose of the scale factor $b$ is ideally to ensure that there is a 1:1 mapping of {\em all} the eigenvalues of $H$ into the $\phi$ = (-1/2,1/2) range. If $b$ is too big, there could be energy eigenvalues that map to phases outside that range. The consequence of this is that these eigenvalues will get mapped back into the (-1/2,1/2) range but at values that, when inverted, yield incorrect energies. This is the phenomenon of aliasing.  In the implementation of Qiskit that we use, a conservative estimate of $b$ is made based on a bound for the largest magnitude eigenvalue of $H$. 
However, the smaller (more conservative) $b$, more binary digits in $\phi$ are required to achieve a given resolution between energy eigenvalues (because the difference in $\phi$ associated with two energies becomes smaller), implying more ancilla are needed. This problem becomes particularly vexing when there are near degenerate states. Thus, in addition to the complication of the lowest eigenenergy not being the most probable (see main text and below), it can be difficult to achieve accurate energies for a small to modest number of ancilla owing to the scale factor, $b$.  
We have found it useful to carry out calculations with larger (less conservative) values of $b$, carefully comparing results to identify any possible aliased energies (only the aliased energies should change with changing $b$ values). The results in Figure 5 of the main text, for example, were obtained with $b$ taken to be nearly six times its default, conservative value, along with experimentation with other $b$ values to ensure that they were not aliased energies.

We also found it instructive to complement some of our QPE calculations by carrying out time propagations of initial fragment wavefunctions, construction of the corresponding autocorrelation functions, and then Fourier analysis via signal processing techniques. This type of approach \cite{somma_quantum_2019} requires only one extra ancilla qubit but requires classical processing, sharing some commonalities with iterative QPE approaches \cite{bauer_quantum_2020, obrien_quantum_2019} . We choose to use as a (short time) signal processing method the Prony method \cite{potts_parameter_2013, obrien_quantum_2019, gray_wave_1992}.
The autocorrelation function $C_k$ is obtained by applying the non-Trotterized unitary
\begin{equation}
    U = e^{i H b \tau}
\end{equation}
to our initial state (here the RHF reference) $k$ = 1, 2, ..., $N$ times, and measuring the overlap between $\langle \Psi_{k\tau} | \Psi_0 \rangle$. This data is then fit to the equation:
\begin{equation}
    C_k = \sum_{s=1}^p h_s e^{i \theta_s k}
\end{equation}
to obtain the $h_s$ and phases $\theta_s$ using Prony's method.  These phases are then converted back to eigenvalues of the Hamiltonian $H$ by:
\begin{equation}
    E_s = \frac{\theta_s}{b \tau}.
\end{equation}
Note that it is easy to see that the $h_s$ correspond to the probabilities of the various eigenvalues.
Using $\tau=0.75$, $p=8$, and $N=20$, we obtain the data in Table \ref{prony} for our transbutadiene system, with R=3.0. Because of the relatively short time propagation,  this approach cannot resolve all the near degeneracies that the more extensive, multi-ancilla QPE calculations in the main text can.  Nonetheless it shows clearly that the initial states in these instances contain significant contributions from excited states and, indeed, the fragment ground states are not even the most populous states. This type of analysis can serve as preliminary information to motivate and guide how the larger scale QPE calculations should be carried out.

\begin{table} [h]
    \centering
    \caption{Coefficients and corresponding energy eigenvalues obtained via Prony analysis of the autocorrelation function.}
    \label{prony}
    \begin{tabular}{ c c | c c }
    \hline
    \hline
    \multicolumn{2}{c|}{Fragment 1} & \multicolumn{2}{c}{Fragment 2} \\
    $h_s$ & $E_s$ & $h_s$ & $E_s$ \\
    \hline
    0.482 & -2.7880 & 0.482 &  -2.7879 \\
    0.210 & -3.3041 & 0.210 & -3.3040 \\
    0.186 & -3.0805 & 0.186 & -3.0797 \\
    0.122 & -1.5885 & 0.122 & -1.5885 \\
    \hline
    \hline
    \end{tabular}
\end{table}

\section*{S2: Ancilla Qubit Estimation}
The number of ancilla qubits required for a QPE calculation can be estimated by~\cite{li_error_2022}:
\begin{equation}
t \approx n + \log \frac{1}{\Delta E^2 r^{2p} \epsilon}
\label{an_q}
\end{equation}
where $1/2^{n}$ is the required precision of the phase while $\Delta E$ is the minimum phase separation. In our case, as we are interested in only the ground state energy, it can be replaced by the difference between the ground and first excited state energies for each system. $r$ corresponds to the number of Trotter steps with $p$ relating to the fidelity with which the unitary can be prepared on the circuit, and $1 - \epsilon$ to the desired success probability, which needs to be high in order for all fragments to be prepared simultaneously for the VQE portion of the algorithm.
We set the number of Trotter steps to 6, 9 and 4 respectively for the (H$_2$)$_2$, (H$_2$)$_4$, and C$_4$H$_6$ systems. $\epsilon$ is set to 0.33, resulting in a success probability of 0.66. We assume $p \approx 1$.

\section*{S3: Ancilla qubit-dependence of Gate Counts}
\begin{figure}[htbp]
\centering
\includegraphics[width=0.6\linewidth]{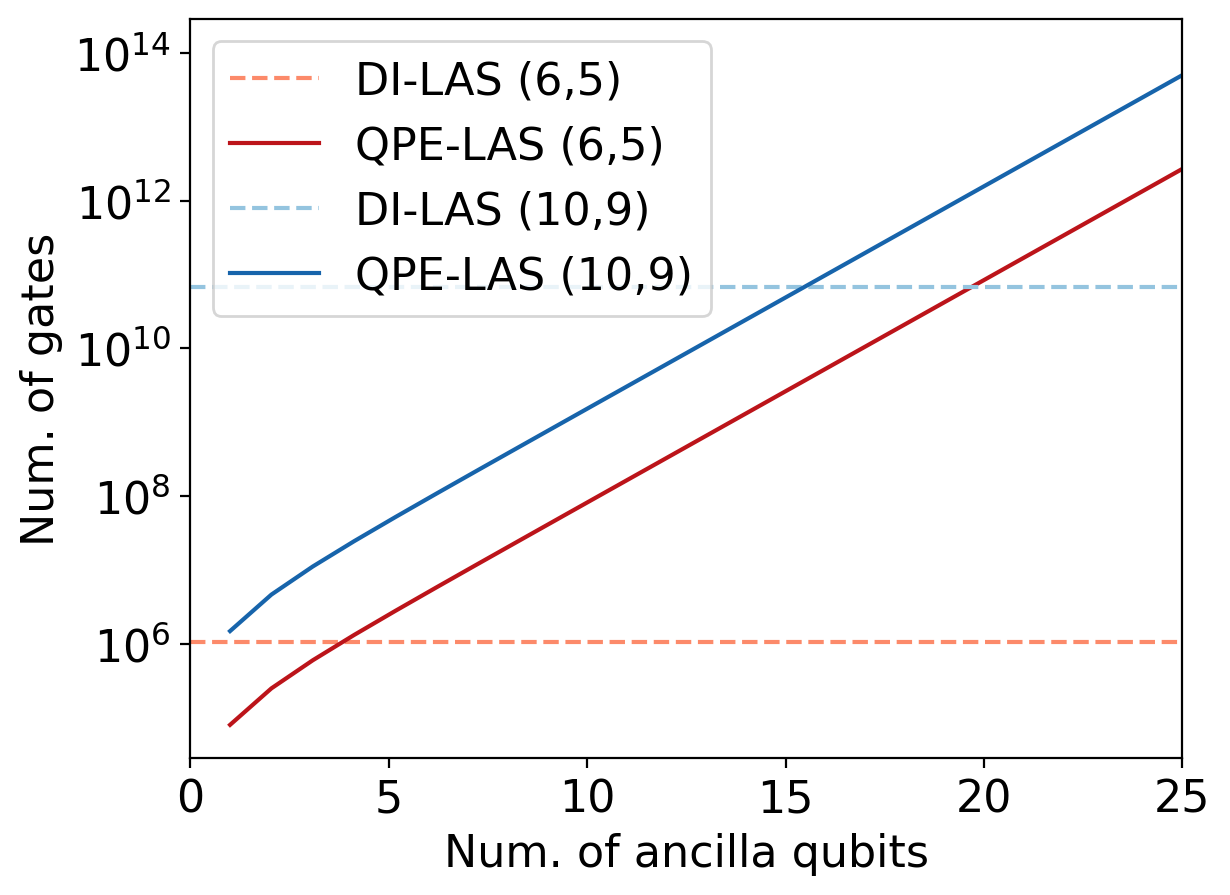}
\caption{Estimated per-fragment gate counts for the state preparation circuits tested as a function of the number of ancilla qubits used for two active space sizes. QPE-LAS-UCC counts are estimated using 10 Trotter repetitions.}
\label{an_counts}
\end{figure}

We use the [Fe(H$_2$O)$_4$]$_2$bpym$^{+4}$ (bpym = 2,2'-bipyrimidine) system and two localized active spaces to study the effect of increasing the number of ancilla qubits on the number of CNOT gates required per fragment in QPE-LAS-UCC. The smaller active space is ((6,5),(6,5)) and the larger active space is ((10,9),(10,9)), both localized to each Fe center. The exponential scaling of the DI-LAS-UCC with the number of active spin orbitals means that at smaller numbers of ancilla qubits, the required number of CNOT gates is orders of magnitude smaller than those required by DI-LAS-UCC. As seen in Figure \ref{an_counts}, which contains estimates of the number of gates required for a given number of ancilla qubits, as the active space grows, the crossover point where it makes sense to use QPE-based method shifts to larger numbers of ancilla qubits. Thus, we can obtain our desired precision without sacrificing efficiency when considering large active spaces.

\section*{S4: Additional data}
\begin{figure}[ht]
\centering
\includegraphics[width=0.6\columnwidth]{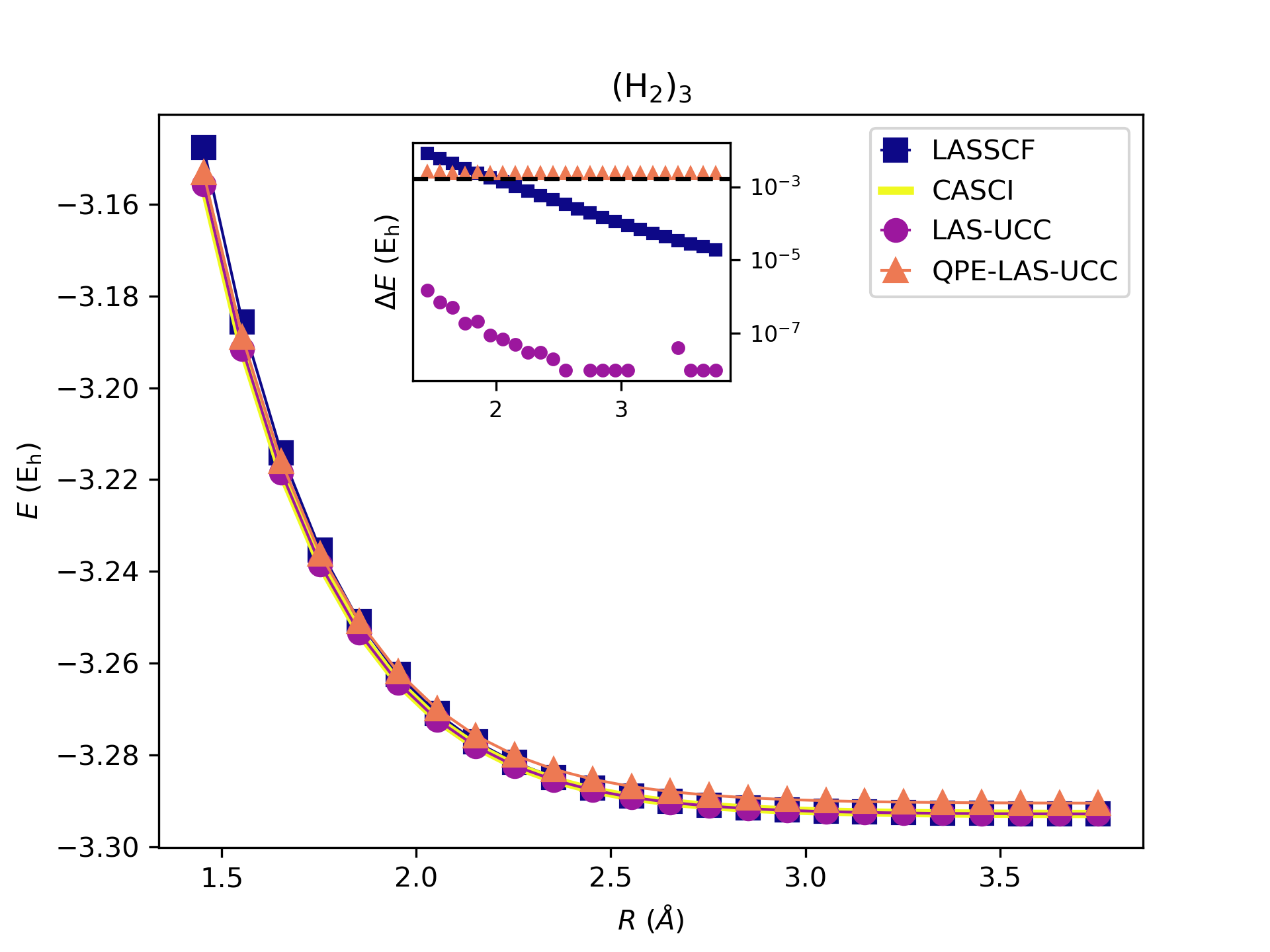}
\caption{VQE energy in Hartrees 
($E_h$) as a function of the distance between molecular centers of mass for (H$_2$)$_3$ as the molecules are moved apart.}
\label{h4_pes}
\end{figure}

\begin{figure}[h]
\begin{subfigure}[b]{0.5\columnwidth}
\includegraphics[width=0.9\linewidth]{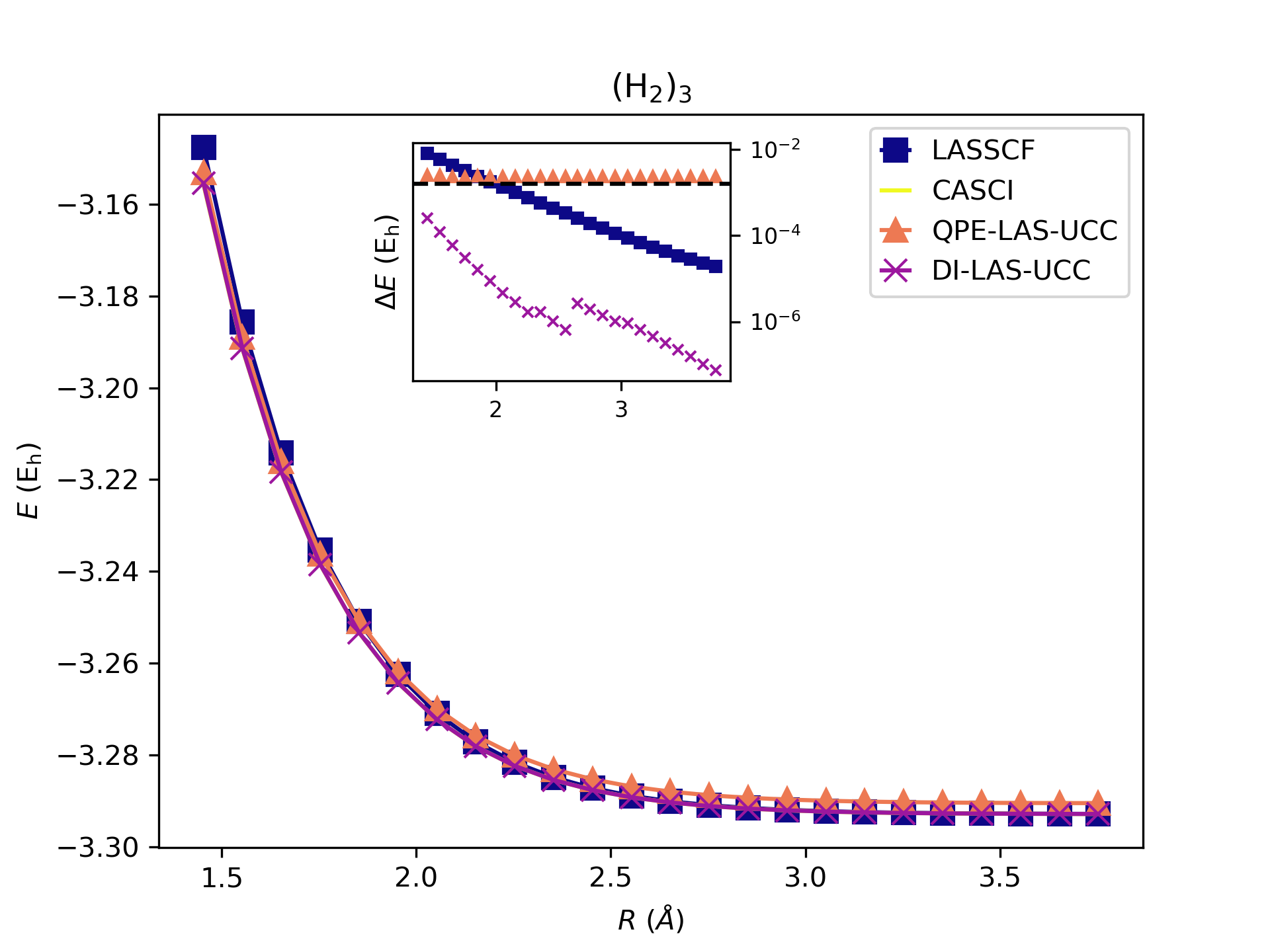}
\caption{}
\label{h4_pes_init}
\end{subfigure}
\begin{subfigure}[b]{0.5\columnwidth}
\includegraphics[width=0.83\linewidth]{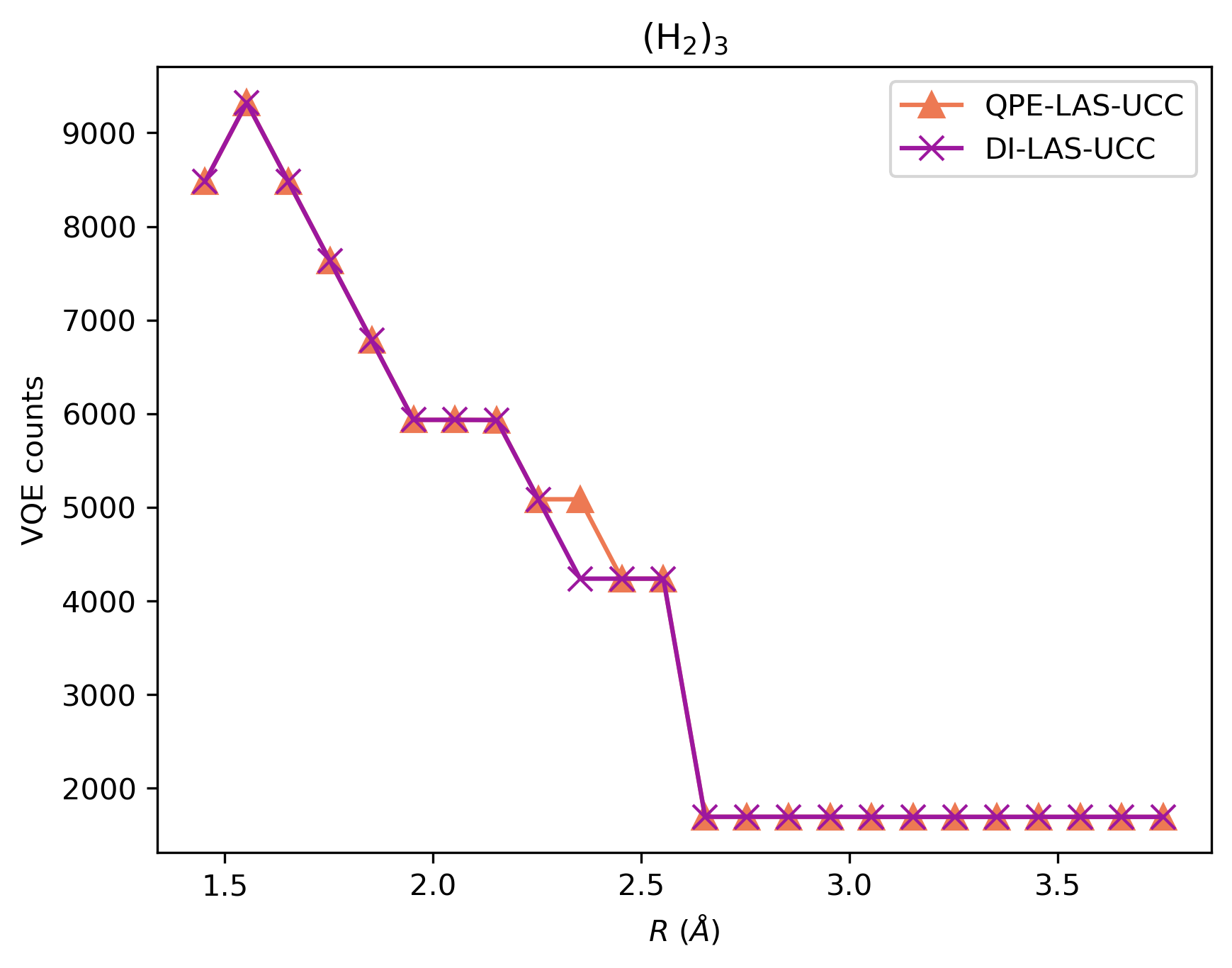}
\caption{}
\label{h4_counts_init}
\end{subfigure}
\caption{(a) VQE energy error with respect to the CASCI reference in Hartrees ($E_h$) and (b) Number of VQE function evaluations, as a function of the distance between molecular centers of mass for (H$_2$)$_3$ as the molecules are moved apart.}
\end{figure}

\begin{table}[h]                                                    
    \centering
    \caption{Resource estimates for QPE and QPE-LAS-UCC for the implementation of a single unitary using a single Trotter step for the [Fe(H$_2$O)$_4$]$_2$bpym$^{+4}$ system.}
    \label{qpe_gates}
    \begin{tabular}{ c r c r } 
    \hline
    \hline
    CAS & QPE CNOTs & LAS & QPE-LAS-UCC CNOTs \\
    \hline
    (12, 10) & 250,644 &  ((6,5),(6,5)) & 16,036  \\
    (14, 12) & 613,912 &  ((8,7),(6,5)) & 50,760  \\
    (16, 14) & 1,324,756 &  ((8,7),(8,7)) & 85,484  \\
    (18, 16) & 2,576,800 &  ((10,9),(8,7)) & 191,800  \\
    (20, 18) & 4,611,108 &  ((10,9),(10,9)) & 298,116  \\
    \hline
    \hline
    \end{tabular}
\end{table}

\clearpage
\section*{S5: Molecular Geometries}
\begin{table}[h]                                                    
    \centering
    \caption{Atomic coordinates of (H$_2$)$_4$ 
 (\AA ngstroms)}
    \label{las_ucc_files/h2_4}
    \begin{tabular}{ c c c c } 
    \hline
    \hline
    Atomic & & & \\
    Symbol & X & Y & Z \\
    \hline
    H &   0.000000 &   0.000000 &   0.000000 \\
    H &   1.000000 &   0.000000 &   0.000000 \\
    H &   0.200000 &   1.600000 &   0.100000 \\
    H &   1.159166 &   1.300000 &  -0.100000 \\
    H &   0.359685 &   3.002880 &   0.049335 \\
    H &   1.358647 &   2.797120 &  -0.049335 \\
    H &   0.624706 &   4.577880 &   0.197837 \\
    H &   1.452792 &   4.122120 &  -0.197837 \\
    \hline
    \hline
    \end{tabular}
\end{table}

\begin{table}[h]
    \centering
    \caption{Atomic coordinates of trans-butadiene
 (\AA ngstroms)}
    \label{las_ucc_files/c4h6}
    \begin{tabular}{ c c c c } 
    \hline
    \hline
    Atomic & & & \\
    Symbol & X & Y & Z \\
    \hline
    C &  -1.833376 &   0.000000 &   0.362184 \\
    H &  -2.781284 &   0.000000 &  -0.139873 \\
    H &  -1.857552 &  -0.000000 &   1.436643 \\
    C &  -0.667088 &   0.000000 &  -0.323584 \\
    H &  -0.680782 &   0.000000 &  -1.400031 \\
    C &   0.667088 &  -0.000000 &   0.323584 \\
    H &   0.680782 &  -0.000000 &   1.400031 \\
    C &   1.833376 &  -0.000000 &  -0.362184 \\
    H &   1.857552 &   0.000000 &  -1.436643 \\
    H &   2.781284 &  -0.000000 &   0.139873 \\
    \hline
    \hline
    \end{tabular}
\end{table}

\begin{table}[h!]
    \centering
    \caption{Atomic coordinates of [Mn(NH$_3$)$_4$]oxamide[Cu(NH$_3$)$_2$]$^{2+}$
 (\AA ngstroms)}
    \label{las_ucc_files/cu_mn_comple}
    \begin{tabular}{ c c c c }
    \hline
    \hline
    Atomic & & & \\
    Symbol & X & Y & Z \\
    \hline
    Cu &  -3.019630 &  -0.007541 &  -0.018036 \\
    Mn &   2.414767 &  -0.023478 &   0.030690 \\
    N &  -1.567623 &   1.238448 &  -0.410772 \\
    N &  -1.580212 &  -1.277148 &   0.270959 \\
    N &  -4.477455 &  -1.377388 &   0.234355 \\
    N &  -4.445900 &   1.415487 &  -0.119243 \\
    C &  -0.396910 &   0.713913 &  -0.231658 \\
    C &  -0.402291 &  -0.751244 &   0.193785 \\
    O &   0.719754 &   1.274368 &  -0.382229 \\
    O &   0.709685 &  -1.305693 &   0.409195 \\
    N &   4.035049 &  -1.399594 &   0.787214 \\
    N &   2.805466 &  -1.226444 &  -1.858758 \\
    N &   2.810332 &   1.223361 &   1.887095 \\
    N &   3.835899 &   1.504798 &  -0.838902 \\
    H &  -5.344417 &  -1.227673 &  -0.285776 \\
    H &  -4.761790 &  -1.469833 &   1.213052 \\
    H &  -4.766482 &   1.576647 &  -1.077764 \\
    H &  -5.296267 &   1.268483 &   0.427975 \\
    H &  -4.167500 &  -2.313239 &  -0.035954 \\
    H &  -1.585917 &  -2.255151 &   0.559517 \\
    H &  -1.551477 &   2.213879 &  -0.705967 \\
    H &  -4.086023 &   2.320991 &   0.189752 \\
    H &   3.778153 &   1.430066 &   2.139229 \\
    H &   2.369771 &   2.137275 &   1.758658 \\
    H &   2.395577 &   0.869641 &   2.752500 \\
    H &   4.689943 &   1.721736 &  -0.321823 \\
    H &   3.329935 &   2.393266 &  -0.885199 \\
    H &   4.156909 &   1.359875 &  -1.798233 \\
    H &   3.771503 &  -1.400275 &  -2.140808 \\
    H &   2.359005 &  -0.860138 &  -2.702860 \\
    H &   2.392501 &  -2.155451 &  -1.746877 \\
    H &   3.745933 &  -2.377479 &   0.704362 \\
    H &   4.295729 &  -1.309775 &   1.771472 \\
    H &   4.925795 &  -1.353260 &   0.288437 \\
    \hline
    \hline
    \end{tabular}
\end{table} 

\begin{table}
    \centering
    \caption{Atomic coordinates of
[Fe(H$_2$O)$_4$]$_2$bpym$^{+4}$ (bpym = 2,2’-bipyrimidine) (\AA ngstroms)}
    \label{las_ucc_files/fe_comple}
    \begin{tabular}{ c c c c }
    \hline
    \hline
    Atomic & & & \\
    Symbol & X & Y & Z \\
    \hline
    Fe &  -3.004847 &   0.000000 &   0.000000 \\
    Fe &   3.004847 &   0.000000 &   0.000000 \\
    C &   0.000000 &   3.409725 &   0.000000 \\
    C &  -1.181171 &   2.688098 &   0.000000 \\
    C &   0.000000 &   0.740752 &   0.000000 \\
    C &   1.181171 &   2.688098 &   0.000000 \\
    C &   0.000000 &  -0.740752 &   0.000000 \\
    C &  -1.181171 &  -2.688098 &   0.000000 \\
    H &  -2.146819 &  -3.183003 &   0.000000 \\
    C &   0.000000 &  -3.409725 &   0.000000 \\
    C &   1.181171 &  -2.688098 &   0.000000 \\
    H &   0.000000 &   4.493448 &   0.000000 \\
    H &  -2.146819 &   3.183003 &   0.000000 \\
    H &   2.146819 &   3.183003 &   0.000000 \\
    H &   0.000000 &  -4.493448 &   0.000000 \\
    H &   2.146819 &  -3.183003 &   0.000000 \\
    N &   1.181856 &  -1.349331 &   0.000000 \\
    N &   1.181856 &   1.349331 &   0.000000 \\
    N &  -1.181856 &   1.349331 &   0.000000 \\
    N &  -1.181856 &  -1.349331 &   0.000000 \\
    O &  -3.144689 &   0.000000 &   2.173794 \\
    H &  -3.215015 &   0.767135 &   2.755999 \\
    H &  -3.215015 &  -0.767135 &   2.755999 \\
    O &  -4.456206 &  -1.575287 &   0.000000 \\
    H &  -4.991908 &  -1.818728 &  -0.767267 \\
    H &  -4.991908 &  -1.818728 &   0.767267 \\
    O &  -4.456206 &   1.575287 &   0.000000 \\
    H &  -4.991908 &   1.818728 &  -0.767267 \\
    H &  -4.991908 &   1.818728 &   0.767267 \\
    O &   4.456206 &  -1.575287 &   0.000000 \\
    H &   4.991908 &  -1.818728 &  -0.767267 \\
    H &   4.991908 &  -1.818728 &   0.767267 \\
    O &   4.456206 &   1.575287 &   0.000000 \\
    H &   4.991908 &   1.818728 &  -0.767267 \\
    H &   4.991908 &   1.818728 &   0.767267 \\
    O &   3.144689 &   0.000000 &   2.173794 \\
    H &   3.215015 &  -0.767135 &   2.755999 \\
    H &   3.215015 &   0.767135 &   2.755999 \\
    O &  -3.144689 &   0.000000 &  -2.173794 \\
    H &  -3.215015 &   0.767135 &  -2.755999 \\
    H &  -3.215015 &  -0.767135 &  -2.755999 \\
    O &   3.144689 &   0.000000 &  -2.173794 \\
    H &   3.215015 &  -0.767135 &  -2.755999 \\
    H &   3.215015 &   0.767135 &  -2.755999 \\
    \hline
    \hline
    \end{tabular}
\end{table}
\clearpage
\bibliographystyle{quantum}
\bibliography{references}